\documentclass{JHEP3}
\usepackage{amssymb,amsmath}
\usepackage{feynmp}
\usepackage{cite}

\newcommand{\bea}{\begin{eqnarray}}
\newcommand{\eea}{\end{eqnarray}}
\newcommand{\bean}{\begin{eqnarray*}}
\newcommand{\eean}{\end{eqnarray*}}

\def\half{\frac{1}{2}}

\def\sqbra#1{\left[ #1\right|}

\def\ket#1{\left| #1\right\rangle}
\def\sqket#1{\left| #1\right]}

\def\gb #1{ \left\langle #1 \right]}
\def\tgb #1{ \left[ #1 \right\rangle}
\def\cb #1{ \left[ #1 \right]}

\def\vev#1{\left\langle #1 \right\rangle}

\def\Label#1{\label{#1}%
  \smash{\hbox to0pt{\raise1ex\hbox{\tiny[#1]}\hss}}}

\title{On-shell recursion for massive fermion currents}
\author{
Ruth Britto and Alexander Ochirov\\
Institut de Physique Th\'eorique, CEA-Saclay, F-91191 Gif-sur-Yvette
cedex, France \\
Email: \email{ruth.britto@cea.fr, alexander.ochirov@cea.fr}
}

\abstract{
We analyze the validity of BCFW recursion relations for currents of $n-2$ gluons and two massive quarks, where one of the quarks is off shell and the remaining particles are on shell.  These currents are gauge-dependent and can be used as ingredients in the unitarity-based approach to computing one-loop amplitudes.  
The validity of BCFW recursion relations is well known to depend on the so-called boundary behavior of the currents as the momentum shift parameter goes to infinity.  With off-shell currents, a new potential problem arises, namely unphysical poles that depend on the choice of gauge.
We identify conditions under which boundary terms are absent 
and unphysical poles are avoided, so that there is a natural recursion relation. 
In particular, we are able to choose a gauge in which we construct a valid shift for currents with at least $n-3$ gluons of the same helicity.  We derive an analytic formula in the case where all gluons have the same helicity.  
As by-products, 
we prove the vanishing boundary behavior of general off-shell objects in Feynman gauge, and
we find a compact generalization of Berends-Giele gluon currents with a generic reference spinor.
}
\preprint{IPhT-t12/079}

\begin{document}

\section{Introduction}

Scattering amplitudes in gauge theories rapidly become difficult to compute as the number of external legs increases. The difficulty is encountered especially when seeking analytic expressions.  However, recent advances involving on-shell techniques have enabled the discovery of many new formulas for amplitudes.  Notably, the BCFW construction generates tree-level amplitudes efficiently and compactly through recursion relations, via the Cauchy residue theorem.

The phrase ``on-shell technique'' refers to methods of computing scattering amplitudes in which certain propagators are taken to their on-shell limits, and amplitudes are then constructed from knowledge of the corresponding factorization properties.  The key is that these factorization limits are calculated in terms of amplitudes of lower complexity, i.e. fewer legs or loops.

It is interesting to study amplitudes (or rather, currents) in which one or more legs is continued {\em off shell}, since they carry even more information than on-shell amplitudes.  For example, the Berends-Giele recursion relation among gluon currents in Yang-Mills theory \cite{Berends:1987me}
is not only computationally powerful for numerical results, but was also the crucial stepping stone to establishing the first formulas for gluon amplitudes with arbitrary numbers of legs, in certain helicity configurations \cite{Berends:1987me, Kosower:1989xy}.  It is still possible to consider the limits in which internal propagators go on-shell and apply the BCFW construction to find recursion relations \cite{Feng:2011twa}.  Compared to the recursion relations for on-shell amplitudes, the ones for currents require committing to a gauge choice, and summing over all internal polarization states, including unphysical polarizations.

In this paper, we seek compact analytic forms for currents of $n-2$ gluons and two massive quarks, where one of the quarks is off shell and the remaining particles are on shell.  These currents are key ingredients of an on-shell method of
computing 1-loop amplitudes with external massive fermions \cite{Britto:2011cr}.  Such amplitudes are of particular interest in the context of LHC searches for new physics, where production of top quarks plays a large role in both signals and backgrounds.  Massive fermion currents can be computed with the off-shell Berends-Giele recursion \cite{Berends:1987me}.  In \cite{Rodrigo:2005eu}, this was used to give a compact result in the case where all gluons have the same helicity, with a particular gauge choice relative to the massive spinors.

We study the validity of the BCFW construction \cite{Britto:2005fq,Badger:2005jv} for these massive fermion currents.  The construction begins by shifting the momenta of a pair of on-shell external legs by $+zq$ and $-zq$ respectively, where $z$ is a complex variable and $q$ is obtained by requiring that both legs remain on shell after the shift.
Then, the residue theorem produces a recursion relation from poles in $z$ taking values where propagators go on shell.  The construction breaks down if there are poles from other sources.  In Yang-Mills theory, the only other possible source is a ``boundary term,'' from a nonvanishing limit as $z$ is taken to infinity.\footnote{If the theory is sufficiently well understood, it is possible to include a boundary term explicitly at each step of the recursion \cite{Feng:2009ei,Feng:2010ku,Benincasa:2011kn,Benincasa:2011pg,Feng:2011twa}.}
For off-shell currents, there is another problematic source of poles, which we call ``unphysical poles.''  They are due to the dependence on gauge choice, and they spoil the recursion relation, since we have no information about how to calculate their residues independently.

We identify conditions under which the boundary terms and unphysical poles vanish for massive fermion currents, so that the BCFW construction produces a recursion relation.  We then proceed to solve the recursion relation in the particular case where all gluons have equal helicities.  Compared to the more compact result of \cite{Rodrigo:2005eu}, our formula also requires all gluons to use the same reference spinor but preserves the genericity of its value.

Our analysis of boundary terms is based on grouping Feynman diagrams conveniently and applying the Ward identity and inductive arguments.  The argument establishes the absence of boundary terms for {\em general} off-shell objects in  Feynman gauge, provided that there are two on-shell gluons available to construct the momentum shift.  

In our study of unphysical poles, we use off-shell gluon currents of the type originally derived by Berends and Giele \cite{Berends:1987me}.  We are motivated to generalize the currents in which one gluon has opposite helicity to all the others, by taking its reference spinor to be arbitrary.  When the opposite-helicity gluon is color-adjacent to the off-shell leg, we find a very compact form for the current.  When it is centrally located among the other gluons, we prove that the current is, in fact, independent of the arbitrary reference spinor.

This paper is organized as follows.  The remainder of this introduction contains details of our conventions and notation.  Section 2 reviews the BCFW construction of recursion relations in the context of massive fermion currents and describes the origin of boundary terms and unphysical poles.  Section 3 derives sufficient conditions for good boundary behavior; Section 4 derives sufficient conditions for the absence of unphysical poles. Section 5 presents sample results from our recursion relations.  In particular, we find a closed form for $n$-point currents in which all gluons have the same helicity.  Section 6 is a summary with proposals for future work.  Appendix A works out a technical point in the proof of Section 3, namely the use of Ward identities in Feynman gauge.  Appendix B presents a generalization of Berends-Giele currents with one gluon of opposite helicity, in which the choice of reference spinor is relaxed.  Appendix C contains a fully non-recursive formula as an alternative to the $n$-point current given in Section 5, and outlines its derivation.

For reference, analytic formulas for on-shell amplitudes of gluons with massive quarks may be found in \cite{Quigley:2005cu, Schwinn:2006ca, Ozeren:2006ft, Ferrario:2006np, Schwinn:2007ee, Hall:2007mz, Chen:2011sba, Boels:2011zz, Huang:2012gs}, all of which use BCFW recursion relations, sometimes in combination with SUSY Ward identities or Berends-Giele recursion.

\subsection{Conventions and notation}

Momenta of gluons are directed outward, while momenta of fermions are directed inward.  We will be considering color-ordered amplitudes and off-shell currents with one massive fermion line, for example, $ i J \left( 1_{\bar{Q}}^*, 2_Q^{}, 3_g, 4_g, \dots, n_g\right)$, where the star means that the indicated leg is considered off-shell, while the remaining legs are on-shell. We do not include the propagator for the off-shell leg in our definition. For this current, the quark line has its arrow pointing from leg 2 to leg 1.  When the quark line matrices are read against the arrow, then the gluon indices are contracted in reverse numerical order.  In slightly different notation, we can write 
      \begin{equation}
            i J \left( 1_{\bar{Q}}^*, 2_Q^{}, 3_g^{h_3}, 4_g^{h_4}, \dots, n_g^{h_n} \right)
            = | n^{h_n} \dots 4^{h_4} 3^{h_3} | 2 ) ,
      \label{altnotation}
	\end{equation}
where the round bracket $|2)$ can be equal to either $ |2 \rangle $ or $|2]$, depending on its spin.  This notation emphasizes the fact that the current is a spinorial object. For example, to obtain the corresponding amplitude, one should first put $p_1$ on shell and then contract the current with either $[1|$ or $ \langle 1| $.

Our color-ordered Feynman rules use the following gluon vertices:
      \begin{eqnarray}
      \parbox{17mm}{
      \begin{fmffile}{graph67} \fmfframe(10,10)(10,0){
      \begin{fmfgraph*}(30,30)
            \fmflabel{$p, \lambda$}{g1}
            \fmflabel{$q, \mu$}{g2}
            \fmflabel{$r, \nu$}{g3}
            \fmftop{g2,g3}
            \fmfbottom{g1}
            \fmf{photon}{g2,v,g3}
            \fmf{photon}{g1,v}
      \end{fmfgraph*} }
      \end{fmffile}
      }
      & = & -\frac{i}{\sqrt{2}} \left[ g_{\lambda \mu} (p - q)_{\nu}
                                     + g_{\mu \nu} (q - r)_{\lambda}
                                     + g_{\nu \lambda} (r - p)_{\mu} \right] ,
      \label{vertex3G} \\
      \parbox{17mm}{
      \begin{fmffile}{graph68} \fmfframe(10,10)(10,0){
      \begin{fmfgraph*}(30,30)
            \fmflabel{$\lambda$}{g1}
            \fmflabel{$\mu$}{g2}
            \fmflabel{$\nu$}{g3}
            \fmflabel{$\rho$}{g4}
            \fmfleft{g1,g2}
            \fmfright{g4,g3}
            \fmf{photon}{g1,v,g2}
            \fmf{photon}{g3,v,g4}
      \end{fmfgraph*} }
      \end{fmffile}
      }
      & = & \frac{i}{2} \left[ 2 g_{\lambda \nu} g_{\mu \rho}
                               - g_{\lambda \rho} g_{\mu \nu}
                               - g_{\lambda \mu} g_{\nu \rho} \right] .
      \label{vertex4G}
	\end{eqnarray}

The polarization vector for a gluon of momentum $p$ is, depending on helicity \cite{Berends:1981rb,DeCausmaecker:1981bg,Kleiss:1985yh,Xu:1986xb,Gunion:1985vca},
\bea
\varepsilon^\mu_{p-} = -\frac{1}{\sqrt{2}}\frac{\tgb{n_p|\gamma^\mu|p}}{\cb{n_p p}},
\qquad
\varepsilon^\mu_{p+} = \frac{1}{\sqrt{2}}\frac{\gb{n_p|\gamma^\mu|p}}{\vev{n_p p}},
\label{eq:pvectors}
\eea
 where $n_p$ is an arbitrary but fixed ``reference'' momentum satisfying $n_p^2=0$ and either $\vev{n_p p} \neq 0$ or $\cb{n_p p} \neq 0$, so that the denominator is nonzero.  The null reference momenta are chosen independently for each gluon.  The set of reference momenta is what we refer to as the gauge choice for a particular calculation, within the Feynman gauge used throughout the spinor-helicity formalism.  Any current we construct with a specific gauge choice is expected to fit into a larger calculation, such as the one-loop computations of \cite{Britto:2011cr}, in which all external legs are on-shell, so that ultimately, after being combined with other ingredients computed in the same gauge, no trace of the gauge choice remains. Therefore the reference spinors can be chosen to maximize computational convenience.  We delay the choice as far as possible, so that convenience can be evaluated later in the full context of a larger calculation.
   
The spinors for the massive fermions satisfy the Dirac equation.  We do not require any further details of their definitions, so any of various conventions (e.g. \cite{Kleiss:1985yh}, \cite{Schwinn:2005pi}) can be used.  
The massless limit is smooth.

We use the Lorentz vector $P_{i,j}$ to denote the sum of color-adjacent momenta in increasing cyclic order, between and including legs $i$ and $j$.

\section{The recursive construction}

We choose to apply the momentum shift to a pair of gluons, since they are always on-shell in this context.  The shift denoted by $\tgb{kl}$ shifts the momenta of gluons labeled by $k$ and $l$ as follows:
\bea
\hat{\ket{k}} = \ket{k}, \qquad
\hat{\sqket{k}} = \sqket{k}-z\sqket{l}, \qquad
\hat{\ket{l}} = \ket{l}+z\ket{k}, \qquad
\hat{\sqket{l}} = \sqket{l},
\label{kl-shift}
\eea
or
\bea
\widehat{p}_k  = p_k - z q, \qquad \widehat{p}_l & = p_l + z q,
\eea
with the complex-valued shift vector
\bea
q^\mu = \half \gb{k|\gamma^\mu|l}.
\label{q}
\eea

The BCFW construction is to
apply the residue theorem on $iJ(z)/z$ to recover the current as 
\bea
iJ(z=0) = - \sum_{\rm{poles}} {\rm{Res}}\left(\frac{iJ(z)}{z}\right).
\eea

To obtain recursion relations among currents, we must have no pole at infinity.  This is assured if $J(z)$ goes to zero in the limit $z \to \infty$, which we call {\em good boundary behavior}.

The remaining poles can have two possible origins, due to their construction from Feynman rules:  (1) the vanishing of a propagator, or (2) the vanishing of the denominator of a polarization vector, when written as in (\ref{eq:pvectors}).  

The first type of pole, from a vanishing propagator, is familiar from the recursion relation for on-shell amplitudes.  The corresponding residues are easy to evaluate as the product of two currents or amplitudes with fewer legs, since the vanishing of the propagator is an on-shell condition.

The second type of pole will be called an {\em unphysical pole}.  In an on-shell amplitude, the reference spinor in such a pole could be freely chosen to eliminate the $z$ dependence in the denominator, but now with a quark line off-shell, we must fix all reference spinors from the start, and they play an explicit role.  These unphysical poles are problematic, since their location has no natural physical meaning, and we have no independent way of computing their residues.  Thus, we will find  conditions on the currents and shift that prevent the appearance of unphysical poles.

With good boundary behavior and no unphysical poles, there will be a recursion relation that takes the schematic form
\bea
i J_n(z=0) = \sum_{k, z_{\rm poles}} \sum_{h} i {\hat J}_{k+1}(z) ~\frac{i}{P^2}~ i {\hat J}_{n-k+1}(z),
\eea
where the $i{\hat J}$  are currents and amplitudes with fewer legs; the hat denotes their evaluation at shifted momentum values; $P$ is the momentum flowing between them which goes on-shell at the pole $z$; and the second sum is over internal helicities.  Of course, the propagator acquires a mass if it is fermionic.  If all of the off-shell legs belong to just one of the two currents on the right-hand side, then the other is replaced by a shifted on-shell amplitude, $i{\hat M}(z)$.

\section{Boundary behavior}

      In this section, we study the behavior as $z \to \infty$ of the  fermionic current $iJ\left( 1_{\bar{Q}}^*, 2_Q^*, 3_g, 4_g, \dots, n_g\right) $ under the  $ [ k l \rangle $ shift (\ref{kl-shift}), where $k$ and $l$ represent any of the gluons. We will conclude that the boundary term vanishes in the helicity cases $(k_g^-,l_g^+)$, $(k_g^-,l_g^-)$ and thus $(k_g^+,l_g^+)$ as well, for a generic gauge choice.  In fact, our argument is much more general, since none of the unshifted gluons need to be on shell. Moreover, the number of fermion lines is not crucial either.

\subsection{Choice of shift: helicities and polarizations}

      Consider the superficial boundary behavior of individual Feynman diagrams, following the flow of the additional momentum $zq$.  Without the polarization vectors, the diagrams where the $zq$ momentum goes only through 3-gluon vertices and gluon propagators behaves the worst --- as $ O(z) $. If $zq$ runs through a 4-gluon vertex or through a fermion line, then such a diagram already behaves as $ O(1) $ or better. Now, if we contract the vector indices for the $\tgb{kl}$-shifted gluons with their polarization vectors,
      \begin{eqnarray} \begin{aligned}
            \hat{\varepsilon}_{k-}^{\mu} & = - \frac{1}{\sqrt{2}}
                  \frac{ [n_k |\gamma^{\mu}| k \rangle }
                       { [n_k k] - z [n_k l] } &
            \hat{\varepsilon}_{k+}^{\mu} & =  \frac{1}{\sqrt{2}}
                  \frac{ \langle n_k| \gamma^{\mu}| k] -z \langle n_k |\gamma^{\mu} |l] }
                       { \langle n_k k \rangle } \\
            \hat{\varepsilon}_{l-}^{\nu} & = - \frac{1}{\sqrt{2}}
                  \frac{ [n_l| \gamma^{\nu} |l \rangle + z [n_l |\gamma^{\nu}| k \rangle }
                       { [n_l l] } \ &
            \hat{\varepsilon}_{l+}^{\nu} & =  \frac{1}{\sqrt{2}}
                  \frac{ \langle n_l |\gamma^{\nu} |l] }
                       { \langle n_l l \rangle + z \langle n_l k \rangle },
      \label{polvectors}
	\end{aligned} \end{eqnarray}
we see that for a generic gauge choice the off-shell current superficially has $ O\left( \frac{1}{z} \right) $ behavior in the $ (k_g^-,l_g^+) $ case, $O(z)$ in the  $ (k_g^-,l_g^-) $ and $ (k_g^+,l_g^+) $ cases, and $O(z^3)$ in the  $ (k_g^+,l_g^-) $ case. Note that this behavior can be altered by special gauge choices, i.e. if $ n_k = l $ or $ n_l = k $, then $ \hat{\varepsilon}_{k-}^{\mu} $ and $ \hat{\varepsilon}_{l+}^{\nu} $ lose their $z$ dependence instead of being $ O\left( \frac{1}{z} \right) $.

      The helicity case $(k_g^-,l_g^+)$ is thus safe automatically, for a generic gauge choice.  In Subsection \ref{subsec:mixhelspecialgauge}, we will discuss boundary behavior for a special gauge choice that will be needed in the following sections.
      
      In the remainder of this section, we will prove that in a generic gauge (where $n_k \neq l$), the off-shell current with helicities $ (k_g^-,l_g^-) $ also vanishes at infinity, at least as $ O\left( \frac{1}{z} \right) $. To do that, let us multiply this current by the $z$-independent factor $ - \sqrt{2} [n_l l] $, so that only the numerator, $ [n_l |\gamma^{\nu}| \hat{l}\rangle $, remains contracted with the $l$-th gluon's Lorentz index $\nu$. The resulting expression depends only linearly on $[n_l|$, which is a 2-dimensional massless spinor and thus can be expressed as a linear combination of any two independent spinors of the same kind:
	\begin{equation}
            [n_l| = \alpha [l| + \beta [n_k| .
      \label{n_l}
      \end{equation}
Therefore it is enough to show that we get $ O\left( \frac{1}{z} \right) $ behavior for the two special cases $n_l = l$ and $n_l=n_k$.  Let us examine them one by one.

\subsection{Like-helicity shift, first term}

      The first term in (\ref{n_l}) yields $ [l|\gamma^{\nu}|\hat{l} \rangle = 2 \hat{l}^{\nu} $, making it possible to use the Ward identity, which diagrammatically can be expressed as follows:
      \begin{equation} \begin{aligned}
      \hat{l}_{\nu}
      \left[ \parbox{37mm}{ \begin{fmffile}{graph0}
      \fmfframe(12,18)(0,18){ \begin{fmfgraph*}(80,60)
            \fmflabel{$1$}{q1}
            \fmflabel{$2$}{q2}
            \fmflabel{$3$}{g3}
            \fmflabel{$\vdots$}{d1}
            \fmflabel{$\hat{k}$}{gk}
            \fmflabel{$\dots$}{d2}
            \fmflabel{$\hat{l},\nu$}{gl}
            \fmflabel{$\vdots$}{d3}
            \fmflabel{$n$}{gn}
            \fmfleft{q2,g3,d1,gk}
            \fmftop{d2}
            \fmfright{q1,gn,d3,gl}
            \fmf{fermion}{c,q1}
            \fmf{fermion}{q2,c}
            \fmf{photon}{g3,c}
            \fmf{photon}{gk,c}
            \fmf{photon}{gl,c}
            \fmf{photon}{gn,c}
            \fmfblob{0.25w}{c}
      \end{fmfgraph*} }
      \end{fmffile} } \right]
      = g \left\{
      \parbox{37mm}{ \begin{fmffile}{graph1}
      \fmfframe(12,18)(0,18){ \begin{fmfgraph*}(80,60)
            \fmflabel{$1-\hat{l}$}{q1}
            \fmflabel{$2$}{q2}
            \fmflabel{$3$}{g3}
            \fmflabel{$\vdots$}{d1}
            \fmflabel{$\hat{k}$}{gk}
            \fmflabel{$\dots$}{d2}
            \fmflabel{$\vdots$}{d3}
            \fmflabel{$n$}{gn}
            \fmfleft{q2,g3,d1,gk}
            \fmftop{d2}
            \fmfright{q1,gn,d3,gl}
            \fmf{fermion}{c,q1}
            \fmf{fermion}{q2,c}
            \fmf{photon}{g3,c}
            \fmf{photon}{gk,c}
            \fmf{phantom}{gl,c}
            \fmf{photon}{gn,c}
            \fmfblob{0.25w}{c}
      \end{fmfgraph*} }
      \end{fmffile} }
      \hspace{2mm} \right.
      & +
      \hspace{2mm}
      \parbox{37mm}{ \begin{fmffile}{graph2}
      \fmfframe(12,18)(0,18){ \begin{fmfgraph*}(80,60)
            \fmflabel{$1$}{q1}
            \fmflabel{$2-\hat{l}$}{q2}
            \fmflabel{$3$}{g3}
            \fmflabel{$\vdots$}{d1}
            \fmflabel{$\hat{k}$}{gk}
            \fmflabel{$\dots$}{d2}
            \fmflabel{$\vdots$}{d3}
            \fmflabel{$n$}{gn}
            \fmfleft{q2,g3,d1,gk}
            \fmftop{d2}
            \fmfright{q1,gn,d3,gl}
            \fmf{fermion}{c,q1}
            \fmf{fermion}{q2,c}
            \fmf{photon}{g3,c}
            \fmf{photon}{gk,c}
            \fmf{phantom}{gl,c}
            \fmf{photon}{gn,c}
            \fmfblob{0.25w}{c}
      \end{fmfgraph*} }
      \end{fmffile} } \\
      + \hspace{4mm}
      \parbox{39mm}{ \begin{fmffile}{graph3}
      \fmfframe(12,18)(0,18){ \begin{fmfgraph*}(80,60)
            \fmflabel{$1$}{q1}
            \fmflabel{$2$}{q2}
            \fmflabel{$3+\hat{l}$}{g3}
            \fmflabel{$\vdots$}{d1}
            \fmflabel{$\hat{k}$}{gk}
            \fmflabel{$\dots$}{d2}
            \fmflabel{$\vdots$}{d3}
            \fmflabel{$n$}{gn}
            \fmfleft{q2,g3,d1,gk}
            \fmftop{d2}
            \fmfright{q1,gn,d3,gl}
            \fmf{fermion}{c,q1}
            \fmf{fermion}{q2,c}
            \fmf{photon}{g3,c}
            \fmf{photon}{gk,c}
            \fmf{phantom}{gl,c}
            \fmf{photon}{gn,c}
            \fmfblob{0.25w}{c}
      \end{fmfgraph*} }
      \end{fmffile} }
      + \dots & + \hspace{2mm} \left.
      \parbox{43mm}{ \begin{fmffile}{graph4}
      \fmfframe(12,18)(0,18){ \begin{fmfgraph*}(80,60)
            \fmflabel{$1$}{q1}
            \fmflabel{$2$}{q2}
            \fmflabel{$3$}{g3}
            \fmflabel{$\vdots$}{d1}
            \fmflabel{$\hat{k}$}{gk}
            \fmflabel{$\dots$}{d2}
            \fmflabel{$\vdots$}{d3}
            \fmflabel{$n+\hat{l}$}{gn}
            \fmfleft{q2,g3,d1,gk}
            \fmftop{d2}
            \fmfright{q1,gn,d3,gl}
            \fmf{fermion}{c,q1}
            \fmf{fermion}{q2,c}
            \fmf{photon}{g3,c}
            \fmf{photon}{gk,c}
            \fmf{phantom}{gl,c}
            \fmf{photon}{gn,c}
            \fmfblob{0.25w}{c}
      \end{fmfgraph*} }
      \end{fmffile} }
      \right\} .
      \label{ward}
      \end{aligned} \end{equation}

      Each diagram on the right-hand side of (\ref{ward}) is supposed to have an appropriate gauge group generator contracted with the leg to which the momentum $ \hat{l} $ is added, but that is irrelevant for the discussion of $ z \rightarrow \infty $ behavior. For any leg that is initially on shell, such as the $k$-th gluon, the corresponding right-hand-side term should naturally vanish, because the resulting leg would go off shell and thus would be left out when extracting the on-shell pole residue according to the standard LSZ procedure.

      Now consider what the Feynman rules tell us about the diagrams on the right-hand side. In case there are off-shell gluons, if a diagram has the $zq$ momentum going from the $k$-th gluon to another through 3-gluon vertices, it must now behave no worse than $ O\left( \frac{1}{z} \right) $! Indeed, it still has $ \hat{\varepsilon}_{k-}^{\mu} \sim O\left( \frac{1}{z} \right) $ on the $k$-th leg, and in addition to that the gluon propagator on the off-shell leg is now $ O\left( \frac{1}{z} \right) $ as well. If $zq$ runs through a 4-gluon vertex or through a fermion line and ends up still on a gluon leg, then such a diagram will behave at most as $  O\left( \frac{1}{z^2} \right) $. A new ingredient here is the diagrams that have $zq$ momentum flowing through 3-point vertices to an off-shell fermion leg, but it is easy to see that they will also behave like $ O\left( \frac{1}{z} \right) $ or better. In sum, applying the Ward identity in this way reduces the maximal superficial power of $z$ at infinity by two.

      There is one technical caveat about this argument: strictly speaking, the Ward identity (\ref{ward}) is valid for a ghostless gauge, whereas in Feynman gauge it is necessary to introduce some extra terms on the right-hand side. We address this issue carefully in Appendix \ref{app:wardid} and find that the argument still holds in Feynman gauge.

\subsection{Like-helicity shift, second term}

      Now we consider the case $ n_l = n_k = n $. It turns out to be possible to deduce some interesting facts about the boundary behavior of an off-shell current simply from the Feynman rules.

\subsubsection{Gluon trees, leading power of $z$}

      We start with the leading $ O(z) $ diagrams, in which the $zq$ momentum flows only through 3-gluon vertices and gluon propagators, which thus behave as $ O(z) $ at most. Let us look closely at the part of such a diagram that is directly adjacent to the $zq$ momentum flow, i.e.\ just a gluon tree with all but the $k$-th and $l$-th legs off-shell and their propagators amputated, keeping in mind that any of the off-shell legs can be extended by any sort of $z$-independent tree, including a fermion line.
      \begin{figure}[h]
      \begin{center}
      \vspace{4mm}
      \begin{fmffile}{graph6}
      \fmfframe(0,0)(0,-5){ \begin{fmfgraph*}(100,50)
            \fmflabel{$\lambda_1$}{g1}
            \fmflabel{$\hat{k}$}{gk}
            \fmflabel{$\lambda_2$}{g2}
            \fmflabel{$\lambda_3$}{g3}
            \fmflabel{$\lambda_4$}{g4}
            \fmflabel{$\hat{l}$}{gl}
            \fmflabel{$\lambda_5$}{g5}
            \fmftop{gk,,g2,g3,g4}
            \fmfbottom{g1,g5,,gl}
            \fmf{photon}{gk,v1,v5,v2,v3,v4,gl}
            \fmffreeze
            \fmf{photon}{g1,v1}
            \fmf{photon}{g2,v2}
            \fmf{photon}{g3,v3}
            \fmf{photon}{g4,v4}
            \fmf{photon}{g5,v5}
      \end{fmfgraph*} }
      \end{fmffile}
      \end{center}
      \caption{A generic gluon tree diagram with only 3-point vertices. \label{gluontree}}
      \end{figure}
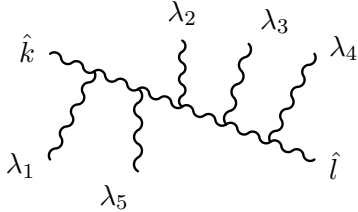
A tree with $n$ legs will have $(n-2)$ vertices $ \sim O(z) $,  $(n-3)$ internal propagators $ \sim O\left( \frac{1}{z} \right) $ and $(n-2)$ free indices. The highest power of $z$ will be accumulated if we pick up $zq^{\lambda}$ from each vertex, a $ z [n |\gamma^{\nu}| k \rangle $ term from the $l$-th gluon's polarization vector, and another $ [n |\gamma^{\mu}| k \rangle $ coming from the $k$-th gluon: $n$ vectors in total. Apart from that, vertices and propagator numerators can only offer various combinations of metric tensors, and the fact that there are only $(n-2)$ free indices means that at least one contraction will take place among those vectors. But any such contraction eliminates a power of $z$, since
      \begin{eqnarray*}
            [n |\gamma^{\nu}| k \rangle q_{\mu} & = 0 , \\
            q^2 & = 0 .
	\end{eqnarray*}
So the leading $ O(z) $ term vanishes and we are left only with $ O(1) $ at most.

\subsubsection{Fermion line insertion, leading power of $z$}

      Similarly, the leading  $ O(1) $ term vanishes for the diagrams in which $zq$ flows through the fermion line. To see this, consider once again only the terms directly adjacent to the $zq$ momentum flow, i.e.\ the relevant part of the fermion line and mostly off-shell gluon trees on both of its sides.
      \begin{figure}[h]
      \begin{center}
      \vspace{4mm}
      \begin{fmffile}{graph7}
      \fmfframe(0,0)(0,-15){ \begin{fmfgraph*}(100,50)
            \fmflabel{$\lambda_1$}{g1}
            \fmflabel{$\hat{k}$}{gk}
            \fmflabel{$\lambda_2$}{g2}
            \fmflabel{$\lambda_3$}{g3}
            \fmflabel{$\lambda_4$}{g4}
            \fmflabel{$\hat{l}$}{gl}
            \fmftop{,g2,g3,}
            \fmfleft{q2,g1,gk}
            \fmfright{q1,g4,gl}
            \fmf{fermion}{q2,vf2,vf3,vf1,q1}
            \fmffreeze
            \fmf{photon}{gk,v1,v2,vf2}
            \fmf{photon}{vf1,v4,gl}
            \fmf{photon}{g1,v1}
            \fmf{photon}{g2,v2}
            \fmf{photon}{g3,vf3}
            \fmf{photon}{g4,v4}
      \end{fmfgraph*} }
      \end{fmffile}
      \end{center}
      \caption{A generic diagram with $zq$ flowing through the fermion line and only 3-point gluon vertices. \label{fermionline}}
      \end{figure}
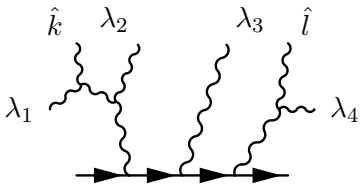

      The leading power of $z$ in a generic diagram with $n$ gluons will now be attained by accumulating $(n-1)$ powers of $zq^{\lambda}$ from 3-gluon vertices and the numerators of fermion propagators. As before, $ [n |\gamma^{\mu}| k \rangle $ and $ z [n |\gamma^{\nu}| k \rangle $ will come from the $k$-th and $l$-th gluons respectively, so in total we have $(n+1)$ vectors with only $(n-2)$ free indices to attribute to them. Note that the relevant part of the fermion line consists of an odd number of $\gamma$-matrices and thus can always be expressed as a linear combination of eight basic matrices $ \{\gamma^{\mu}\}_{\mu=0}^3 \cup \{\gamma^{\mu} \gamma^5\}_{\mu=0}^3 $ just by using the standard formula:
	\begin{equation} \begin{aligned}
            \gamma^{\lambda} \gamma^{\mu} \gamma^{\nu}
            = g^{\lambda \mu} \gamma^{\nu}
            - g^{\lambda \nu} \gamma^{\mu}
            + g^{\mu \nu} \gamma^{\lambda}
            + i \epsilon^{\lambda \mu \nu \rho} \gamma_{\rho} \gamma^5 .
      \label{threegamma}
	\end{aligned} \end{equation}
The free index of the $ \gamma^{\mu} $ or $ \gamma^{\mu} \gamma^5 $ can either be left free as an off-shell gluon index (leaving us with only $(n-3)$ free indices left for $(n+1)$ vectors) or be contracted with one of the $(n+1)$ vectors. So the number of free indices is smaller than the number of vectors at least by two. The difference with the previous case is that now we have not only metric tensors to do the index-contraction work, but also the totally antisymmetric tensor coming from (\ref{threegamma}). So, lowering the number of free vector indices by two can be achieved by either dotting one vector to another, in which case we get zero just as in the gluon-tree case; or by contracting three vectors to one antisymmetric tensor constructing terms like $ \epsilon_{\lambda \mu \nu \rho}  \cdot z q^{\lambda} \cdot [n |\gamma^{\mu}| k \rangle \cdot z [n| \gamma^{\nu}| k \rangle $, all of which vanish due to the fact that we have copies of only two vectors in the leading $O(z)$ term. Thus all diagrams with $zq$ momentum flow through a fermion line necessarily vanish at $ z \rightarrow \infty $.

\subsubsection{Gluon trees, next-to-leading power of $z$}

      What is left to consider is the possible $O(1)$ contribution from gluon trees. To begin with, calculate the simplest gluon tree, i.e.\ a single 3-gluon vertex contracted with the two polarization vectors as given in (\ref{polvectors}):
      \begin{equation} \begin{aligned}
      \parbox{17mm}{
      \begin{fmffile}{graph8} \fmfframe(0,0)(0,0){
      \begin{fmfgraph*}(30,30)
            \fmflabel{$\lambda$}{g}
            \fmflabel{$\hat{k}$}{gk}
            \fmflabel{$\hat{l}$}{gl}
            \fmftop{gk,gl}
            \fmfbottom{g}
            \fmf{photon}{gk,v,gl}
            \fmf{photon}{g,v}
      \end{fmfgraph*} }
      \end{fmffile}
      }
      & = \hat{\varepsilon}_{k-}^{\mu}
          \left( g_{\mu \nu} (\hat{k} - \hat{l})_{\lambda}
               + g_{\nu \lambda} (2 \hat{l} + \hat{k})_{\mu}
               - g_{\lambda \mu} (2 \hat{k} + \hat{l})_{\nu}
          \right)
          \hat{\varepsilon}_{l-}^{\nu} \\
      & = - \frac{1}{ 2 z [nl]^2 }
          \left( z \langle k |\gamma_{\lambda} |n] \cdot 2 \langle k | l | n] )
               + \langle k |\gamma_{\lambda}| n] \langle l |\gamma^{\nu} |n]
                 \cdot z \langle k |\gamma_{\nu}| l]
          \right) + O\left( \frac{1}{z} \right) \\
      & = - \frac{\langle k |\gamma_{\lambda} |n]}{2 [nl]^2 }
          \left( 2 \langle k l \rangle [ln]
               + 2 \langle l k \rangle [ln]
          \right) + O\left( \frac{1}{z} \right)  = O\left( \frac{1}{z} \right) .
      \label{gluonthree}
      \end{aligned} \end{equation}
Here the $O(z)$ terms vanish trivially in accord with our previous considerations, but we see from the Fierz identity that the $O(1)$ term is canceled as well. At four legs these cancellations continue to take place, but start to involve $O(1)$ diagrams with a single quartic vertex insertion, for instance:
      \begin{eqnarray} \begin{aligned}
      \parbox{17mm}{
      \begin{fmffile}{graph12} \fmfframe(10,10)(10,0){
      \begin{fmfgraph*}(30,40)
            \fmflabel{$\lambda_1$}{g1}
            \fmflabel{$\lambda_2$}{g2}
            \fmflabel{$\hat{k}$}{gk}
            \fmflabel{$\hat{l}$}{gl}
            \fmfleft{g1,gk}
            \fmfright{gl,g2}
            \fmf{photon}{gk,v2,v1,gl}
            \fmf{photon}{g1,v1}
            \fmf{photon}{g2,v2}
      \end{fmfgraph*} }
      \end{fmffile}
      }
      +
      \parbox{22mm}{
      \begin{fmffile}{graph13} \fmfframe(10,10)(10,0){
      \begin{fmfgraph*}(40,30)
            \fmflabel{$\lambda_1$}{g1}
            \fmflabel{$\lambda_2$}{g2}
            \fmflabel{$\hat{k}$}{gk}
            \fmflabel{$\hat{l}$}{gl}
            \fmfleft{g1,gk}
            \fmfright{gl,g2}
            \fmf{photon}{gk,v1,v2,gl}
            \fmf{photon}{g1,v1}
            \fmf{photon}{g2,v2}
      \end{fmfgraph*} }
      \end{fmffile}
      }
      +
      \parbox{17mm}{
      \begin{fmffile}{graph14} \fmfframe(10,10)(10,0){
      \begin{fmfgraph*}(30,30)
            \fmflabel{$\lambda_1$}{g1}
            \fmflabel{$\lambda_2$}{g2}
            \fmflabel{$\hat{k}$}{gk}
            \fmflabel{$\hat{l}$}{gl}
            \fmfleft{g1,gk}
            \fmfright{gl,g2}
            \fmf{photon}{gk,v,gl}
            \fmf{photon}{g1,v}
            \fmf{photon}{g2,v}
      \end{fmfgraph*} }
      \end{fmffile}
      }
      & = O\left( \frac{1}{z} \right) .
      \label{gluonfour}
      \end{aligned} \end{eqnarray}

\begin{figure}[h]
      \begin{center}
      \parbox{17mm}{
      \begin{fmffile}{graph15} \fmfframe(0,0)(0,-5){
      \begin{fmfgraph*}(40,60)
            \fmflabel{$\hat{k}$}{gk}
            \fmflabel{$\hat{l}$}{gl}
            \fmflabel{$p'$}{out}
            \fmflabel{$r'$}{in}
            \fmftop{gk,gl}
            \fmfbottom{out,in}
            \fmf{photon}{gk,v,gl}
            \fmf{photon}{g,v}
            \fmf{fermion}{in,g,out}
      \end{fmfgraph*} }
      \end{fmffile}
      }
      + \hspace{5mm}
      \parbox{25mm}{
      \begin{fmffile}{graph16} \fmfframe(0,0)(0,-5){
      \begin{fmfgraph*}(60,40)
            \fmflabel{$\hat{k}$}{gk}
            \fmflabel{$\hat{l}$}{gl}
            \fmflabel{$p'$}{out}
            \fmflabel{$r'$}{in}
            \fmftop{gk,gl}
            \fmfbottom{out,in}
            \fmf{photon}{gk,vk}
            \fmf{photon}{gl,vl}
            \fmf{fermion}{in,vl,vk,out}
      \end{fmfgraph*} }
      \end{fmffile}
      }
      \end{center}
      \caption{Diagrams for the amplitude of 2 gluon and 1 fermion line. \label{gluonthreeappend}}
      \end{figure}
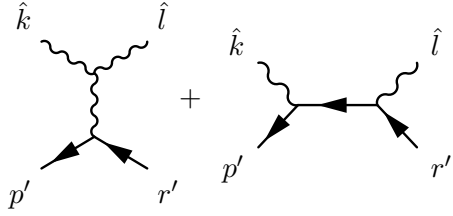

      Evidently, such an intricate cancellation cannot be deduced by examining Feynman diagrams separately. Let us look again at the 3-gluon vertex (\ref{gluonthree}) from another point of view. Attaching a gluon propagator to the off-shell line obviously does not change the power of $z$, and the resulting off-shell 3-gluon current is a Lorentz vector. If we contract it with a simple fermion line, we obtain the first diagram in Fig. \ref{gluonthreeappend}, which is a part of a scattering amplitude --- a gauge invariant object that is well established to behave as $ O\left( \frac{1}{z} \right) $ for the $ (k_g^-,l_g^-) $ shift.

      Moreover, the second diagram in Fig. \ref{gluonthreeappend} has $zq$ momentum flow through its fermion line, so according to our previous discussion it is of order $ O\left( \frac{1}{z} \right) $ by itself. Thus we can conclude that the first one is $ O\left( \frac{1}{z} \right) $ as well. We obtained it by contracting the initial off-shell current vector with a correctly defined fermion line. The freedom of choosing the on-shell fermion momenta and helicities spans the whole Minkowski space.  Therefore, the vector must have the same boundary behavior.

      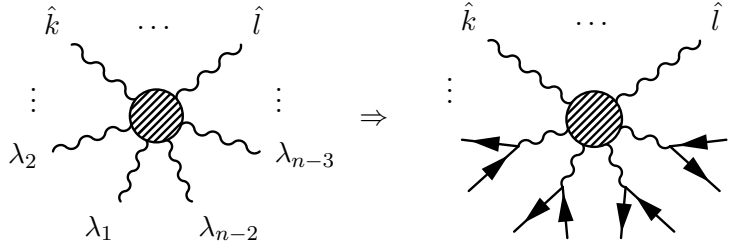
\begin{figure}[h]
      \begin{center}
      \parbox{44mm}{ \begin{fmffile}{graph17}
      \fmfframe(12,5)(0,0){ \begin{fmfgraph*}(80,60)
            \fmflabel{$\lambda_1$}{g1}
            \fmflabel{$\lambda_2$}{g2}
            \fmflabel{$\vdots$}{d1}
            \fmflabel{$\hat{k}$}{gk}
            \fmflabel{$\dots$}{d2}
            \fmflabel{$\hat{l}$}{gl}
            \fmflabel{$\vdots$}{d3}
            \fmflabel{$\lambda_{n-3}$}{gn3}
            \fmflabel{$\lambda_{n-2}$}{gn2}
            \fmfbottom{,g1,gn2,}
            \fmfleft{,g2,d1,gk}
            \fmftop{d2}
            \fmfright{,gn3,d3,gl}
            \fmf{photon}{g1,c}
            \fmf{photon}{g2,c}
            \fmf{photon,tension=1.7}{gk,c}
            \fmf{photon,tension=1.7}{gl,c}
            \fmf{photon}{gn3,c}
            \fmf{photon}{gn2,c}
            \fmfblob{0.25w}{c}
      \end{fmfgraph*} }
      \end{fmffile} }
      $ \Rightarrow $ \hspace{3mm}
      \parbox{48mm}{ \begin{fmffile}{graph18}
      \fmfframe(12,5)(0,-12){ \begin{fmfgraph*}(100,75)
            \fmflabel{$\vdots$}{d1}
            \fmflabel{$\hat{k}$}{gk}
            \fmflabel{$\dots$}{d2}
            \fmflabel{$\hat{l}$}{gl}
            \fmflabel{$\vdots$}{d3}
            \fmfbottom{,out1,in1,outn2,inn2,}
            \fmfleft{,in2,out2,d1,gk}
            \fmftop{d2}
            \fmfright{,outn3,inn3,d3,gl}
            \fmf{photon,tension=1.4}{g1,c}
            \fmf{photon,tension=1.4}{g2,c}
            \fmf{photon,tension=1.7}{gk,c}
            \fmf{photon,tension=1.7}{gl,c}
            \fmf{photon,tension=1.4}{gn3,c}
            \fmf{photon,tension=1.4}{gn2,c}
            \fmf{fermion}{in1,g1,out1}
            \fmf{fermion}{in2,g2,out2}
            \fmf{fermion}{inn3,gn3,outn3}
            \fmf{fermion}{inn2,gn2,outn2}
            \fmfblob{0.21w}{c}
      \end{fmfgraph*} }
      \end{fmffile} }
      \end{center}
      \caption{Contracting a gluon off-shell current with fermion lines. \label{gluontensor}}
      \end{figure}

      Along the same lines, we can now prove a very general statement: \emph{An $n$-gluon off-shell current with only two shifted like-helicity legs on shell behaves as $ O\left( \frac{1}{z} \right) $.} The current has a free Lorentz index for each off-shell leg, so it is actually a tensor of rank $(n-2)$. If we contract every index with its own fermion line (independent of $z$), we will obtain an expression corresponding to a scalar amplitude (Fig. \ref{gluontensor}) which we know behaves as a whole as $ O\left( \frac{1}{z} \right) $ under the $ (k_g^-,l_g^-) $ shift. The freedom of choice of fermion momenta and helicities guarantees that if the contracted expression vanishes at $ z \rightarrow \infty $, then the initial tensor is bound to vanish too.

      \begin{figure}[h]
      \begin{center}
      \parbox{48mm}{ \begin{fmffile}{graph19}
      \fmfframe(0,10)(0,-10){ \begin{fmfgraph*}(125,75)
            \fmflabel{$\hat{k}$}{gk}
            \fmflabel{$\hat{l}$}{gl}
            \fmfbottom{out1,in1,out4,,in4,}
            \fmftop{gk,in2,out2,gl,}
            \fmfright{,out3,in3,}
            \fmf{photon}{g1,gk}
            \fmf{photon}{g1,v1}
            \fmf{photon}{g1,g2}
            \fmf{photon}{g2,v2}
            \fmf{photon}{g2,v5}
            \fmf{photon}{g3,gl}
            \fmf{photon}{g3,v3}
            \fmf{photon}{g3,v4}
            \fmf{fermion}{in1,v1,out1}
            \fmf{fermion}{in2,v2,out2}
            \fmf{fermion}{in3,v3,out3}
            \fmf{fermion}{in4,v4,v5,out4}
      \end{fmfgraph*} }
      \end{fmffile} }
      \hspace{3mm}
      \parbox{60mm}{ \begin{fmffile}{graph20}
      \fmfframe(0,10)(0,-10){ \begin{fmfgraph*}(150,75)
            \fmflabel{$\hat{k}$}{gk}
            \fmflabel{$\hat{l}$}{gl}
            \fmfbottom{out1,,in1,out4,,,in4,}
            \fmftop{gk,in2,out2,gl,}
            \fmfright{,out3,in3,}
            \fmf{photon,tension=0.5}{gk,v5}
            \fmf{photon,tension=1}{g2,v1}
            \fmf{photon,tension=2}{g2,v2}
            \fmf{photon,tension=1}{g2,v7}
            \fmf{photon,tension=0}{gl,v6}
            \fmf{photon}{v3,v4}
            \fmf{fermion}{in1,v1,v5,out1}
            \fmf{fermion}{in2,v2,out2}
            \fmf{fermion}{in3,v3,out3}
            \fmf{fermion}{in4,v4,v6,v7,out4}
      \end{fmfgraph*} }
      \end{fmffile} }
      \end{center}
      \caption{Diagrams for $ 4 \times \{ q \bar{q} \} \rightarrow \hat{g_k} \hat{g_l} $ with $zq$ momentum flow through fermion propagators. \label{gluontensorlacking1}}
      \end{figure}
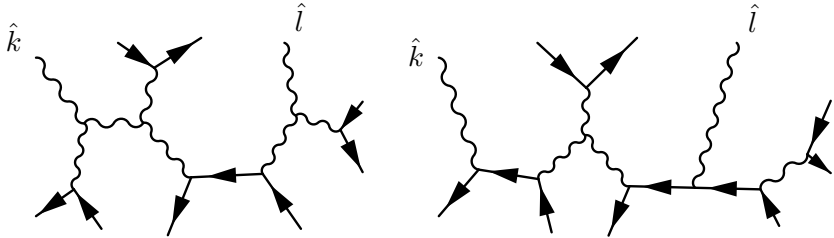

      Of course, we will still lack some diagrams to build the full amplitude, but the lacking terms will in fact be those for which we have already proven the good behavior at $ z \rightarrow \infty $. Indeed, the result resembles an amplitude for a process where $(n-2)$ (distinct) quark-antiquark pairs go to 2 gluons, so it should contain not only the diagrams which are given by the right-hand side of Fig. \ref{gluontensor}, but also those where some fermion lines have multiple fermion vertices and thus have fermion propagator insertions in them. Some of them look like the diagrams which are shown in Fig. \ref{gluontensorlacking1}, i.e.\ have $zq$ flow through at least one of those fermion lines and thus vanish at $ z \rightarrow \infty $.

      \begin{figure}[h]
      \begin{center}
      \parbox{48mm}{ \begin{fmffile}{graph21}
      \fmfframe(0,10)(0,-10){ \begin{fmfgraph*}(125,75)
            \fmflabel{$\hat{k}$}{gk}
            \fmflabel{$\hat{l}$}{gl}
            \fmfbottom{out1,in1,out4,,in4,}
            \fmftop{gk,in2,out2,gl,}
            \fmfright{,out3,in3,}
            \fmf{photon}{g1,gk}
            \fmf{photon,tension=3}{g1,v1}
            \fmf{photon}{g1,g3}
            \fmf{photon,tension=3}{g3,v5}
            \fmf{photon,tension=4}{g3,g2}
            \fmf{photon,tension=2}{g2,v2}
            \fmf{photon}{g2,gl}
            \fmf{photon}{v4,v3}
            \fmf{fermion}{in1,v1,out1}
            \fmf{fermion}{in2,v2,out2}
            \fmf{fermion}{in3,v3,out3}
            \fmf{fermion}{in4,v4,v5}
            \fmf{fermion,tension=3}{v5,out4}
      \end{fmfgraph*} }
      \end{fmffile} }
      \hspace{3mm}
      \parbox{48mm}{ \begin{fmffile}{graph22}
      \fmfframe(0,10)(0,-10){ \begin{fmfgraph*}(125,75)
            \fmflabel{$\hat{k}$}{gk}
            \fmflabel{$\hat{l}$}{gl}
            \fmfleft{,out1,in1,}
            \fmfbottom{,out4,,dummy,,in4,}
            \fmftop{,,gk,,in2,out2,,}
            \fmfright{,out3,in3,gl,,}
            \fmf{photon}{v1,v6}
            \fmf{photon,tension=3}{g1,v5}
            \fmf{photon}{g1,gk}
            \fmf{photon,tension=3}{g1,g2}
            \fmf{photon,tension=2}{g2,gl}
            \fmf{photon,tension=3}{g2,v2}
            \fmf{photon,tension=2}{v3,v4}
            \fmf{fermion}{in1,v1,out1}
            \fmf{fermion}{in2,v2,out2}
            \fmf{fermion}{in3,v3,out3}
            \fmf{fermion}{in4,v4}
            \fmf{fermion,tension=2}{v4,v5,v6}
            \fmf{fermion}{v6,out4}
            \fmf{phantom,tension=4}{v5,dummy}
      \end{fmfgraph*} }
      \end{fmffile} }
      \end{center}
      \caption{Diagrams for $ 4 \times \{ q \bar{q} \} \rightarrow \hat{g_k} \hat{g_l} $, that can be reduced to the case of smaller number of gluon legs. \label{gluontensorlacking2}}
      \end{figure}
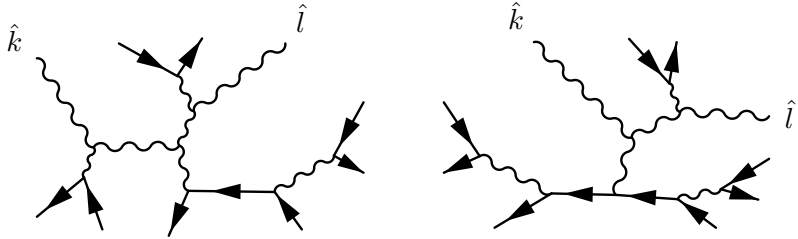

      Others, however, may look like the diagrams shown in Fig. \ref{gluontensorlacking2}, i.~e. contain some fermion lines connected to the shifted gluons only through their connection to other fermion lines. These diagrams can be reduced to the case of a smaller number of off-shell legs in the initial gluon current. 
      Thus, we can construct an inductive argument, for which we have already verified the base case of $n=3$, to see that all the diagrams that we need to add to the contracted $n$-gluon current to form an amplitude behave as $ O\left( \frac{1}{z} \right) $ and the current itself is therefore bound to be $ O\left( \frac{1}{z} \right) $.

      By the way, this inductive proof did not use the weaker $O(1)$ statement of section 3.3.1, though we relied heavily on the $ O\left( \frac{1}{z} \right) $ statement of section 3.3.2. To conclude, let us recall the steps of our argument:

\begin{enumerate}

\item Any diagram with $zq$ momentum flow through at least one fermion propagator behaves well.

\item The boundary behavior of the diagrams with $zq$ momentum flow strictly through gluon propagators is the same as that of a gluon-only off-shell current.

\item Any off-shell current with 3 gluon legs vanishes as $ z \rightarrow \infty $ due to the cancellation which ensures the good behavior of the amplitude $ q \bar{q} \rightarrow \hat{g} \hat{g} $.

\item Any off-shell current with $n$ gluon legs vanishes as $ z \rightarrow \infty $ to ensure the good behavior of the amplitude $ (n-2) \times \{ q \bar{q} \} \rightarrow \hat{g} \hat{g} $, provided the good behavior of the $(n-1)$--gluon current and the diagrams with fermion propagator insertions.

\end{enumerate}

\subsection{Mixed-helicity shift in special gauges}
\label{subsec:mixhelspecialgauge}

      For a generic gauge, the boundary behavior of individual Feynman diagrams under the $ [ k l \rangle $ shift is automatically $ O\left( \frac{1}{z} \right) $ in the mixed-helicity case $(k_g^-,l_g^+)$. But in the following section, we will find that in order to avoid unphysical poles we need to use special gauge choices $ n_k = p_l $ or $ n_l = p_k $. Such gauges eliminate the $z$-dependence from one of the polarization vector denominators and thus turn the superficial behavior into $O(1)$. However, we can easily rephrase our power-counting arguments from the sections 3.3.1 and 3.3.2 for the mixed-helicity case and find that the leading power of $z$ must always involve at least one contraction of two of the following three vectors: $ zq^{\lambda} $, $ [n_k |\gamma^{\mu}| k \rangle $ and $ \langle n_l| \gamma^{\nu} |l] $, with either $g_{\mu \nu}$ or $\epsilon_{\lambda \mu \nu \rho}$. Either by imposing $ n_k = p_l $ or $ n_l = p_k $, we can guarantee that any such contraction will give zero and thus ensure vanishing of the boundary term.

      It is worth noting that if we take $ n_k = p_l $ and $ n_l = p_k $ simultaneously, the superficial boundary behavior is worsened by two powers of $z$, and the argument will in general be invalid. Suppose that we first impose $ n_l = p_k $ and have $n_k$ unfixed. Then we will be guaranteed to have no pole at infinity, but might still have an unphysical pole at $ z_k = [n_k k]/[n_k l] $. Now if we take $ n_k = p_l $, we can see that the pole $z_k$ goes smoothly to infinity. In this way the unphysical pole and the boundary term can be traded one for another, and the problem is to find gauges in which neither survives.

\section{Avoiding unphysical poles}

      In this section, we address the question of unphysical poles, i.e.\ the poles that come from polarization vectors (\ref{polvectors}) instead of propagators. We construct explicit recursive proofs of the vanishing of the unphysical poles for the following currents:

\begin{enumerate}

\item $ [ 3 4 \rangle $-shifted $ i J \left( 1_{\bar{Q}}^*, 2_Q^{}, \hat{3}_g^-, \hat{4}_g^-, 5_g^-, \dots, n_g^- \right) $ with $ n_3 = n_4 = \dots = n_n $ ;

\item $ [ 4 3 \rangle $-shifted $ i J \left( 1_{\bar{Q}}^*, 2_Q^{}, \hat{3}_g^+, \hat{4}_g^-, 5_g^-, \dots, n_g^- \right) $ with $ n_4 = n_5 = \dots = n_n = p_3 $ ;

\item $ [ 3 4 \rangle $-shifted $ i J \left( 1_{\bar{Q}}^*, 2_Q^*, \hat{3}_g^-, \hat{4}_g^+, 5_g^-, \dots, n_g^- \right) $ with $ n_3 = n_5 = \dots = n_n = p_4 $ .

\end{enumerate}

      It is straightforward to prove analogous statements for the currents with the opposite quark off shell or for flipped helicity assignments.  For example, $ i J ( 1_{\bar{Q}}, 2_Q^*, 3_g^-, \dots, \widehat{(n\!-\!1)}_g^-, \hat{n}_g^+ ) $ has no unphysical poles under the $ [ n\!-\!1 | n \rangle $-shift if $ n_3 = \dots = n_{n-1} = p_n $. One can also make a simultaneous flip of all gluon helicities trivially.

      In short, the good gauge choices are:
\begin{itemize}
      \item in the all-minus case, put all reference momenta equal to each other: $   n_i = q $ ;
      \item in the one-plus cases, put reference momenta of negative-helicity gluons equal to the momentum of the positive-helicity gluon: $ n_- = p_+ $.
\end{itemize}

      Note that in the one-plus case with the positive-helicity gluon in central position the unphysical poles vanish for a matrix-valued current $ | n^- \dots 5^- 4^+ 3^- | $ with \emph{both} fermions off shell, i.~e. lacking spinors on both sides of the quark line. In fact, there is strong evidence (see the numerical results in the following section) that it will continue to be true for such one-plus currents $ | n^- \dots  (m\!+\!1)^- m^+ (m\!-\!1)^- \dots 3^- | $ irrespective of the position of the positive-helicity gluon with respect to the fermions.

      In each case, our recursive argument is based on the following expansion \cite{Berends:1987me}.   Consider constructing the $n$-particle current with one fermion line by attaching the $n$-th gluon to the corresponding $(n\!-\!1)$-particle current. Due to color-ordering, it can be coupled directly to the off-shell quark, to the $(n\!-\!1)$-th gluon, or to some more complicated gluon tree. If we focus our attention on those gluon trees that include the $n$-th gluon and attach to the quark line as a whole, we can expand the current according to the number of legs in such trees, as shown pictorially in equation (\ref{vertexrecursion}).
      \begin{equation} \begin{aligned}
      \parbox{30mm}{ \begin{fmffile}{graph41}
      \fmfframe(12,12)(12,0){ \begin{fmfgraph*}(60,60)
            \fmflabel{$1$}{q1}
            \fmflabel{$2$}{q2}
            \fmflabel{$\hat{3}$}{g3}
            \fmflabel{$\hat{4}$}{g4}
            \fmflabel{$5$}{g5}
            \fmflabel{$\dots$}{d}
            \fmflabel{$n$}{gn}
            \fmftop{,g4,,g5,d,}
            \fmfbottom{,q2,,,q1,}
            \fmfleft{,,g3,}
            \fmfright{,,gn,}
            \fmf{fermion,tension=1.5}{q2,c,q1}
            \fmf{photon}{g3,c}
            \fmf{photon}{g4,c}
            \fmf{photon}{g5,c}
            \fmf{photon}{gn,c}
            \fmfblob{0.25w}{c}
      \end{fmfgraph*} }
      \end{fmffile} }
      = &
      \parbox{36mm}{ \begin{fmffile}{graph42}
      \fmfframe(12,12)(12,0){ \begin{fmfgraph*}(75,60)
            \fmflabel{$1$}{q1}
            \fmflabel{$2$}{q2}
            \fmflabel{$\hat{3}$}{g3}
            \fmflabel{$\hat{4}$}{g4}
            \fmflabel{$\dots$}{d}
            \fmflabel{$n\!-\!1$}{gn1}
            \fmflabel{$n$}{gn}
            \fmftop{,g4,,d,gn1,,,,,}
            \fmfbottom{q2,q1}
            \fmfleft{,,,g3,,}
            \fmfright{,,,gn,}
            \fmf{fermion,tension=2.3}{q2,c}
            \fmf{fermion,tension=1.7}{c,v,q1}
            \fmf{photon,tension=0.5}{g3,c}
            \fmf{photon,tension=0.5}{g4,c}
            \fmf{photon,tension=0.5}{gn1,c}
            \fmf{photon,tension=1.5}{gn,v}
            \fmfblob{0.20w}{c}
      \end{fmfgraph*} }
      \end{fmffile} }
      +
      \parbox{40mm}{ \begin{fmffile}{graph43}
      \fmfframe(12,12)(12,0){ \begin{fmfgraph*}(90,60)
            \fmflabel{$1$}{q1}
            \fmflabel{$2$}{q2}
            \fmflabel{$\hat{3}$}{g3}
            \fmflabel{$\hat{4}$}{g4}
            \fmflabel{$\dots$}{d}
            \fmflabel{$n\!-\!2$}{gn2}
            \fmflabel{$n\!-\!1$}{gn1}
            \fmflabel{$n$}{gn}
            \fmftop{,g4,,d,gn2,,gn1,,,}
            \fmfbottom{q2,,,q1,}
            \fmfleft{,,g3,}
            \fmfright{,,,gn,,}
            \fmf{fermion,tension=2.3}{q2,c}
            \fmf{fermion,tension=1.7}{c,v}
            \fmf{fermion,tension=1.2}{v,q1}
            \fmf{photon,tension=0.5}{g3,c}
            \fmf{photon,tension=0.5}{g4,c}
            \fmf{photon,tension=0.5}{gn2,c}
            \fmf{photon,tension=2.0}{v,vg}
            \fmf{photon,tension=1.5}{gn1,vg}
            \fmf{photon,tension=1.5}{gn,vg}
            \fmfblob{0.25h}{c}
      \end{fmfgraph*} }
      \end{fmffile} } \\
      + &
      \parbox{42mm}{ \begin{fmffile}{graph44}
      \fmfframe(12,12)(12,0){ \begin{fmfgraph*}(90,60)
            \fmflabel{$1$}{q1}
            \fmflabel{$2$}{q2}
            \fmflabel{$\hat{3}$}{g3}
            \fmflabel{$\hat{4}$}{g4}
            \fmflabel{$\dots$}{d}
            \fmflabel{$n\!-\!3$}{gn3}
            \fmflabel{$n\!-\!2$}{gn2}
            \fmflabel{$n\!-\!1$}{gn1}
            \fmflabel{$n$}{gn}
            \fmftop{,g4,,d,gn3,,gn2,,,}
            \fmfbottom{q2,,,q1,}
            \fmfleft{,,g3,}
            \fmfright{,,,gn,,gn1,}
            \fmf{fermion,tension=2.3}{q2,c}
            \fmf{fermion,tension=1.7}{c,v}
            \fmf{fermion,tension=1.2}{v,q1}
            \fmf{photon,tension=0.5}{g3,c}
            \fmf{photon,tension=0.5}{g4,c}
            \fmf{photon,tension=0.5}{gn3,c}
            \fmf{photon,tension=2.0}{v,vg}
            \fmf{photon,tension=1.0}{gn2,vg}
            \fmf{photon,tension=1.0}{gn1,vg}
            \fmf{photon,tension=1.0}{gn,vg}
            \fmfblob{0.25h}{vg}
            \fmfblob{0.25h}{c}
      \end{fmfgraph*} }
      \end{fmffile} }
      + \dots +
      \parbox{36mm}{ \begin{fmffile}{graph45}
      \fmfframe(12,12)(12,0){ \begin{fmfgraph*}(90,60)
            \fmflabel{$1$}{q1}
            \fmflabel{$2$}{q2}
            \fmflabel{$\hat{3}$}{g3}
            \fmflabel{$\hat{4}$}{g4}
            \fmflabel{$5$}{g5}
            \fmflabel{$6$}{g6}
            \fmflabel{$\dots$}{d}
            \fmflabel{$n$}{gn}
            \fmftop{,,,g4,,,g5,,g6,}
            \fmfbottom{q2,,,q1,}
            \fmfleft{,,g3,}
            \fmfright{,,,gn,d,}
            \fmf{fermion,tension=2.3}{q2,c}
            \fmf{fermion,tension=1.7}{c,v}
            \fmf{fermion,tension=1.2}{v,q1}
            \fmf{photon,tension=0.75}{g3,c}
            \fmf{photon,tension=0.75}{g4,c}
            \fmf{photon,tension=2.0}{v,vg}
            \fmf{photon,tension=0.75}{g5,vg}
            \fmf{photon,tension=0.75}{g6,vg}
            \fmf{photon,tension=1.0}{gn,vg}
            \fmfblob{0.25h}{vg}
            \fmfblob{0.25h}{c}
      \end{fmfgraph*} }
      \end{fmffile} } \\
      + &
      \parbox{36mm}{ \begin{fmffile}{graph46}
      \fmfframe(12,12)(12,0){ \begin{fmfgraph*}(75,60)
            \fmflabel{$1$}{q1}
            \fmflabel{$2$}{q2}
            \fmflabel{$\hat{3}$}{g3}
            \fmflabel{$\hat{4}$}{g4}
            \fmflabel{$5$}{g5}
            \fmflabel{$\dots$}{d}
            \fmflabel{$n$}{gn}
            \fmftop{,,g4,,g5,}
            \fmfbottom{q2,,q1,}
            \fmfleft{,,g3,}
            \fmfright{,,,gn,d,}
            \fmf{fermion,tension=1.7}{q2,v2,v1}
            \fmf{fermion,tension=1.2}{v1,q1}
            \fmf{photon,tension=1.2}{v1,c}
            \fmf{photon,tension=1.0}{g3,v2}
            \fmf{photon,tension=0.5}{g4,c}
            \fmf{photon,tension=0.5}{g5,c}
            \fmf{photon,tension=0.5}{gn,c}
            \fmfblob{0.20w}{c}
      \end{fmfgraph*} }
      \end{fmffile} }
      +
      \parbox{30mm}{ \begin{fmffile}{graph47}
      \fmfframe(12,12)(12,0){ \begin{fmfgraph*}(60,60)
            \fmflabel{$1$}{q1}
            \fmflabel{$2$}{q2}
            \fmflabel{$\hat{3}$}{g3}
            \fmflabel{$\hat{4}$}{g4}
            \fmflabel{$5$}{g5}
            \fmflabel{$\dots$}{d}
            \fmflabel{$n$}{gn}
            \fmftop{,g4,,g5,d,}
            \fmfbottom{,q2,,q1,}
            \fmfleft{,,g3,}
            \fmfright{,,gn,}
            \fmf{fermion,tension=1.7}{q2,v,q1}
            \fmf{photon,tension=2}{v,c}
            \fmf{photon}{g3,c}
            \fmf{photon}{g4,c}
            \fmf{photon}{g5,c}
            \fmf{photon}{gn,c}
            \fmfblob{0.25w}{c}
      \end{fmfgraph*} }
      \end{fmffile} }
      \label{vertexrecursion} .
      \end{aligned} \end{equation}

\subsection{All-minus currents}

      Let us prove that for the $ [ 3 4 \rangle $-shifted all-minus current $ i J \left( 1_{\bar{Q}}^*, 2_Q^{}, \hat{3}_g^-, \hat{4}_g^-, 5_g^-, \dots, n_g^- \right) $ the residue at the unphysical pole $ z_3 = [n_3 3]/[n_3 4] $ vanishes when we make all the gluon reference momenta equal: $ n_3 = n_4 = \dots = n_n \equiv q $. 
      
      Now if we already know that the residue at the unphysical pole $ z_3 = [q3]/[q4] $ vanishes for all the corresponding off-shell currents with fewer legs, then only the last two diagrams in (\ref{vertexrecursion}) remain to be calculated. That can easily be done just by using color-ordered Feynman rules,
where we make use of the Berends-Giele formula \cite{Berends:1987me} for currents of like-helicity gluons, 
which in our conventions is given by
      \begin{equation}
            i J^{\mu} (1^-,2^-, \dots, n^-) =
            - \frac{[q|\gamma^{\mu} {\not}P_{1,n} |q]}{\sqrt{2} [q 1] [1 2] \dots [n\!-\!1~n] [n q] } .
      \label{berendsgieleminus}
      \end{equation}
Evaluating the sum at the pole $z_3=0$, defined by $ [q \hat{3}] = 0 $, and performing some manipulations using a Schouten identity between the two contributions, we find the 
      residue of (\ref{vertexrecursion}) at the unphysical pole:
      \begin{equation} \begin{aligned}
      \left[q \hat{3}\right] \left\{
      \parbox{40mm}{ \begin{fmffile}{graph50}
      \fmfframe(20,12)(10,12){ \begin{fmfgraph*}(75,60)
            \fmflabel{$1^*$}{q1}
            \fmflabel{$2$}{q2}
            \fmflabel{$\hat{3}^-$}{g3}
            \fmflabel{$\hat{4}^-$}{g4}
            \fmflabel{$5^-$}{g5}
            \fmflabel{$\dots$}{d}
            \fmflabel{$n^-$}{gn}
            \fmftop{,,g4,,g5,}
            \fmfbottom{q2,,q1,}
            \fmfleft{,,g3,}
            \fmfright{,,,gn,d,}
            \fmf{fermion,tension=1.7}{q2,v2,v1}
            \fmf{fermion,tension=1.2}{v1,q1}
            \fmf{photon,tension=1.2}{v1,c}
            \fmf{photon,tension=1.0}{g3,v2}
            \fmf{photon,tension=0.5}{g4,c}
            \fmf{photon,tension=0.5}{g5,c}
            \fmf{photon,tension=0.5}{gn,c}
            \fmfblob{0.20w}{c}
      \end{fmfgraph*} }
      \end{fmffile} }
      +
      \parbox{33mm}{ \begin{fmffile}{graph51}
      \fmfframe(15,12)(10,12){ \begin{fmfgraph*}(60,60)
            \fmflabel{$1^*$}{q1}
            \fmflabel{$2$}{q2}
            \fmflabel{$\hat{3}^-$}{g3}
            \fmflabel{$\hat{4}^-$}{g4}
            \fmflabel{$5^-$}{g5}
            \fmflabel{$\dots$}{d}
            \fmflabel{$n^-$}{gn}
            \fmftop{,g4,,g5,d,}
            \fmfbottom{,q2,,q1,}
            \fmfleft{,,g3,}
            \fmfright{,,gn,}
            \fmf{fermion,tension=1.7}{q2,v,q1}
            \fmf{photon,tension=2}{v,c}
            \fmf{photon}{g3,c}
            \fmf{photon}{g4,c}
            \fmf{photon}{g5,c}
            \fmf{photon}{gn,c}
            \fmfblob{0.25w}{c}
      \end{fmfgraph*} }
      \end{fmffile} }
      \right\} =
      \frac{ i |q] \langle 3|{\not}P_{3,n}|q] [q| ({\not}p_2 - m)  |2) }
           { \langle 3|2|q] [3 4] [4 5] \dots [n\!-\!1~n] [n q] } .
      \label{minusn1}
      \end{aligned} \end{equation}
It becomes obvious that (\ref{minusn1}) vanishes due to the presence of the on-shell spinor $|2)$ next to $({\not}p_2 - m)$.

      To conclude the proof, we do not even need to calculate the base of the recursion separately, because all the preceding formulas were general enough be valid for $  i J \left( 1_{\bar{Q}}^*, 2_Q^{}, \hat{3}_g^-, \hat{4}_g^- \right) $ as well. Indeed, in that case the last two diagrams in (\ref{vertexrecursion}) turn out to be the usual Feynman diagrams with the Berends-Giele current representing just the polarization vector of the shifted $4$th gluon.

\subsection{Currents with a single positive-helicity gluon in extreme position}

      For the $ [ 4 3 \rangle $-shifted current $ i J \left( 1_{\bar{Q}}^*, 2_Q^{}, \hat{3}_g^+, \hat{4}_g^-, 5_g^-, \dots, n_g^- \right) $ we put $ n_4 = \dots = n_n = p_3 $ and rename $ n_3 \equiv q $. Consider the same expansion (\ref{vertexrecursion}). As in the previous case, for a recursive proof of the vanishing of the residue at $ z_3 = - \langle q 3 \rangle / \langle q 4 \rangle $ we only need to calculate the last two diagrams in (\ref{vertexrecursion}).
      We use the following formula for the one-plus Berends-Giele current:
      \begin{equation}
            i J^{\mu} (1^+,2^-, \dots, n^-) =
            - \frac{[1|\gamma^{\mu} {\not}P_{1,n} |1]}{\sqrt{2} [1 2] [2 3] \dots [n\!-\!1~n] [n 1] }
            \left\{ \sum_{l=3}^{n} \frac{ [1| {\not}P_{1,l} {\not}P_{1,l-1} |1] }
                                        { P_{1,l}^2 P_{1,l-1}^2 }
                                 + \frac{ \langle 2q \rangle }{ \langle 21 \rangle \langle 1q \rangle }
            \right\} ,
      \label{berendsgieleplus1}
      \end{equation}
in which we retain dependence on the reference momentum $ n_1 \equiv q $ of the positive-helicity gluon. This generalization is discussed in Appendix \ref{app:berendsgiele}. It turns out that relaxing one reference momentum results in only one extra term in (\ref{berendsgieleplus1}), which subsequently generates the pole for $i {\not}J ( \hat{3}^+, \hat{4}^-, \dots, n^-)$ at $ \langle \hat{3} q \rangle = 0 $.

Using the currents (\ref{berendsgieleminus}) and (\ref{berendsgieleplus1}),
we see again that the residue at $z_3$ vanishes due to the presence of the on-shell spinor $|2)$ next to $({\not}p_2 - m)$:
      \begin{equation} \begin{aligned}
      \langle q \hat{3} \rangle \left\{
      \parbox{40mm}{ \begin{fmffile}{graph56}
      \fmfframe(20,12)(10,12){ \begin{fmfgraph*}(75,60)
            \fmflabel{$1^*$}{q1}
            \fmflabel{$2$}{q2}
            \fmflabel{$\hat{3}^+$}{g3}
            \fmflabel{$\hat{4}^-$}{g4}
            \fmflabel{$5^-$}{g5}
            \fmflabel{$\dots$}{d}
            \fmflabel{$n^-$}{gn}
            \fmftop{,,g4,,g5,}
            \fmfbottom{q2,,q1,}
            \fmfleft{,,g3,}
            \fmfright{,,,gn,d,}
            \fmf{fermion,tension=1.7}{q2,v2,v1}
            \fmf{fermion,tension=1.2}{v1,q1}
            \fmf{photon,tension=1.2}{v1,c}
            \fmf{photon,tension=1.0}{g3,v2}
            \fmf{photon,tension=0.5}{g4,c}
            \fmf{photon,tension=0.5}{g5,c}
            \fmf{photon,tension=0.5}{gn,c}
            \fmfblob{0.20w}{c}
      \end{fmfgraph*} }
      \end{fmffile} }
      +
      \parbox{33mm}{ \begin{fmffile}{graph57}
      \fmfframe(15,12)(10,12){ \begin{fmfgraph*}(60,60)
            \fmflabel{$1^*$}{q1}
            \fmflabel{$2$}{q2}
            \fmflabel{$\hat{3}^+$}{g3}
            \fmflabel{$\hat{4}^-$}{g4}
            \fmflabel{$5^-$}{g5}
            \fmflabel{$\dots$}{d}
            \fmflabel{$n^-$}{gn}
            \fmftop{,g4,,g5,d,}
            \fmfbottom{,q2,,q1,}
            \fmfleft{,,g3,}
            \fmfright{,,gn,}
            \fmf{fermion,tension=1.7}{q2,v,q1}
            \fmf{photon,tension=2}{v,c}
            \fmf{photon}{g3,c}
            \fmf{photon}{g4,c}
            \fmf{photon}{g5,c}
            \fmf{photon}{gn,c}
            \fmfblob{0.25w}{c}
      \end{fmfgraph*} }
      \end{fmffile} }
      \right\} =
      \frac{-i |3] \langle 4q \rangle  \langle q|{\not}P_{3,n}|3] [3| ({\not}p_2 - m)  |2) }
           { \langle q|2|3] \langle 43 \rangle [3 4] [\hat{4} 5] \dots [n\!-\!1~n] [n 3] } .
      \label{plusn1}
      \end{aligned} \end{equation}

      This evaluation is valid as well for $ i J \left( 1_{\bar{Q}}^*, 2_Q^{}, \hat{3}_g^+, \hat{4}_g^- \right) $, thus ensuring the base of the recursive argument.

\subsection{Currents with a single positive-helicity gluon in next-to-extreme position}

      In this section we consider a matrix-valued $\tgb{34}$-shifted current $ i J \left( 1_{\bar{Q}}^*, 2_Q^*, \hat{3}_g^-, \hat{4}_g^+, 5_g^-, \dots, n_g^- \right) $ with the positive-helicity gluon separated from the fermion line by one negative-helicity gluon. We put $ n_3 = n_5 = \dots = n_n = p_4 $ and $ n_4 \equiv q $. In expansion (\ref{vertexrecursion}), we now need to examine the last three terms.  More specifically, we need to examine only their $q$-dependent parts, since these are the ones that can have the unphysical pole at $\vev{q \hat{4}}=0.$ 

      The very last diagram in (\ref{vertexrecursion}) contains the gluon current $ i J \left( \hat{3}^-, \hat{4}^+, 5^-, \dots, n^- \right), $ for which we do not know a simple analytic formula. Fortunately, according to the inductive argument outlined in Appendix \ref{app:berendsgiele}, it does not depend on $q$, so that diagram cannot contribute to the residue at $ \langle q \hat{4} \rangle  = 0 $. We are thus left with the other two diagrams. 
      As before,
      \begin{equation} \begin{aligned}
      \parbox{33mm}{ \begin{fmffile}{graph59}
      \fmfframe(10,12)(10,12){ \begin{fmfgraph*}(75,60)
            \fmflabel{$1^*$}{q1}
            \fmflabel{$2^*$}{q2}
            \fmflabel{$\hat{3}^-$}{g3}
            \fmflabel{$\hat{4}^+$}{g4}
            \fmflabel{$5^-$}{g5}
            \fmflabel{$\dots$}{d}
            \fmflabel{$n^-$}{gn}
            \fmftop{,,g4,,g5,}
            \fmfbottom{q2,,q1,}
            \fmfleft{,,g3,}
            \fmfright{,,,gn,d,}
            \fmf{fermion,tension=1.7}{q2,v2,v1}
            \fmf{fermion,tension=1.2}{v1,q1}
            \fmf{photon,tension=1.2}{v1,c}
            \fmf{photon,tension=1.0}{g3,v2}
            \fmf{photon,tension=0.5}{g4,c}
            \fmf{photon,tension=0.5}{g5,c}
            \fmf{photon,tension=0.5}{gn,c}
            \fmfblob{0.20w}{c}
      \end{fmfgraph*} }
      \end{fmffile} }
      =
      \frac{i}{\sqrt{2} [43]} i {\not}J (\hat{4}^+, 5^-, \dots, n^-)
      \frac{ {\not}p_2 - {\not}\widehat{p}_3 + m }{ (p_2 - \widehat{p}_3)^2 - m^2 }
      \Big( |4] \langle 3| + |3 \rangle [4| \Big) .
      \label{plus43}
      \end{aligned} \end{equation}

We use the formula
      \begin{equation} \begin{aligned}
            i J \left( 1_{\bar{Q}}^*, 2_Q^*, 3_g^-, 4_g^+ \right)
            = \frac{i}{ [n_3 3] \langle n_4 4 \rangle } \Bigg\{
              \Big( |4] \langle n_4| \!+\! |n_4 \rangle [4| \Big)
              \frac{ {\not}p_2 \!-\! {\not}p_3 \!+\! m }{ (p_2 \!-\! p_3)^2 \!-\! m^2 }
            & \Big( |n_3] \langle 3| \!+\! |3 \rangle [n_3| \Big) \\
            + \frac{ [n_3 4] \langle n_4 3 \rangle }{ 2 [3 4] \langle 4 3 \rangle }
                                                       ({\not}p_3 \!-\! {\not}p_4)
            + \frac{ [n_3 4] }{ [3 4] } \Big( |4] \langle n_4| \!+\! |n_4 \rangle [4| \Big)
            - \frac{ \langle n_4 3 \rangle }{ \langle 4 3 \rangle }
                                      & \Big( |n_3] \langle 3| \!+\! |3 \rangle [n_3| \Big)
              \Bigg\} ,
      \end{aligned} \end{equation}
which is derived simply from Feynman rules, to evaluate the third-to-last diagram in expansion (\ref{vertexrecursion}):
      \begin{equation} \begin{aligned}
      \parbox{40mm}{ \begin{fmffile}{graph58}
      \fmfframe(12,12)(12,0){ \begin{fmfgraph*}(90,60)
            \fmflabel{$1^*$}{q1}
            \fmflabel{$2^*$}{q2}
            \fmflabel{$\hat{3}^-$}{g3}
            \fmflabel{$\hat{4}^+$}{g4}
            \fmflabel{$5^-$}{g5}
            \fmflabel{$6^-$}{g6}
            \fmflabel{$\dots$}{d}
            \fmflabel{$n^-$}{gn}
            \fmftop{,,,g4,,,g5,,g6,}
            \fmfbottom{q2,,,q1,}
            \fmfleft{,,g3,}
            \fmfright{,,,gn,d,}
            \fmf{fermion,tension=2.3}{q2,c}
            \fmf{fermion,tension=1.7}{c,v}
            \fmf{fermion,tension=1.2}{v,q1}
            \fmf{photon,tension=0.75}{g3,c}
            \fmf{photon,tension=0.75}{g4,c}
            \fmf{photon,tension=2.0}{v,vg}
            \fmf{photon,tension=0.75}{g5,vg}
            \fmf{photon,tension=0.75}{g6,vg}
            \fmf{photon,tension=1.0}{gn,vg}
            \fmfblob{0.25h}{vg}
            \fmfblob{0.25h}{c}
      \end{fmfgraph*} }
      \end{fmffile} }
      = & -
      \frac{i}{\sqrt{2} \langle q \hat{4} \rangle [43]} i {\not}J (5^-, 6^-, \dots, n^-)
      \frac{ {\not}p_2 - {\not}p_3 - {\not}p_4 + m }{ (p_2 - p_3 - p_4)^2 - m^2 } \\ &
      \Bigg\{
            \Big( |4] \langle q| \!+\! |q \rangle [4| \Big)
            \frac{ {\not}p_2 - {\not}\widehat{p}_3 + m }{ (p_2 - \widehat{p}_3)^2 - m^2 }
          - \frac{ \langle q 3 \rangle }{ \langle 4 3 \rangle }
      \Bigg\}
      \Big( |4] \langle 3| + |3 \rangle [4| \Big) ,
      \label{plus45}
      \end{aligned} \end{equation}

      After using the formulas (\ref{berendsgieleminus}) and (\ref{berendsgieleplus1}) for the Berends-Giele currents and substituting
      \begin{equation}
            |q \rangle = |\hat{4} \rangle \frac{\langle 3q \rangle}{\langle 34 \rangle} 
      \label{g4plus}
      \end{equation}
in the residue of the unphysical pole,
we can combine the $q$-dependent terms of (\ref{plus43}) and (\ref{plus45}) into one term with the following spinor matrix in the middle:
      \begin{equation}
            \frac{ {\not}p_2 - {\not}\widehat{p}_3 - {\not}\widehat{p}_4 + m }
                 { (p_2 - \widehat{p}_3 - \widehat{p}_4)^2 - m^2 } ~ {\not}\widehat{p}_4 ~
            \frac{ {\not}p_2 - {\not}\widehat{p}_3 + m }
                 { (p_2 - \widehat{p}_3)^2 - m^2 } -
            \frac{ {\not}p_2 - {\not}\widehat{p}_3 - {\not}\widehat{p}_4 + m }
                 { (p_2 - \widehat{p}_3 - \widehat{p}_4)^2 - m^2 } +
            \frac{ {\not}p_2 - {\not}\widehat{p}_3 + m }
                 { (p_2 - \widehat{p}_3)^2 - m^2 } = 0 .
      \label{g4plusmatrix}
      \end{equation}

      Having established the induction, we turn back to $ i J \left( 1_{\bar{Q}}^*, 2_Q^*, \hat{3}_g^-, \hat{4}_g^+, 5_g^- \right) $ and see that its expansion (\ref{vertexrecursion}) contains precisely the three diagrams that we have just examined in a more general case. This provides the base of our inductive argument for $ i J \left( 1_{\bar{Q}}^*, 2_Q^*, 3_g^-, 4_g^+, 5_g^-, \dots, n_g^- \right) $.

      Our numerical results (see below) indicate that the statement about the matrix-valued one-plus currents might be true irrespective of the position of the positive-helicity gluon as long as it is separated from the fermion by at least one negative-helicity gluon. Unfortunately, it remains a challenge to show it. Here, we used an explicit formula for $ i J \left( 1_{\bar{Q}}^*, 2_Q^*, 3_g^-, 4_g^+ \right) $ to evaluate one of the diagrams. To prove the vanishing of the unphysical pole for $ i J \left( 1_{\bar{Q}}^*, 2_Q^*, 3_g^-, \dots, (m\!-\!1)_g^-, m_g^+, (m\!+\!1)_g^-, \dots, n_g^- \right) $ in the same manner, one would need to have an explicit formula either for $ i J \left( 1_{\bar{Q}}^*, 2_Q^*, 3_g^-, \dots, (m\!-\!1)_g^-, m_g^+\right) $ or for $ i J \left( 1_{\bar{Q}}^*, 2_Q^*, m_g^+, (m\!+\!1)_g^-, \dots, n_g^- \right) $.

\section{Results for currents}

In this section, we apply the constructions established in the previous section to compute massive fermion currents from recursion relations.  First, we list 3- and 4-point currents as a starting point.  Next, we give a closed-form expression for currents with an arbitrary number of gluons if their helicities are all alike.  A fully non-recursive version and its derivation are given in Appendix \ref{app:closedform}.  Finally, we state our numerical results for shifts producing recursion relations in the case of one gluon of opposite helicity.

\subsection{3-point and 4-point currents}

For completeness, we begin by listing the 3- and 4-point currents, which are straightforward to derive from Feynman rules, with full freedom of the choice of reference spinors.

     \begin{subequations} \begin{align}
            i J \left( 1_{\bar{Q}}^*, 2_Q^*, 3_g^- \right) = - i &
            \frac{ |n_3] \langle 3| \!+\! |3 \rangle [n_3| }
                 { [n_3 3] } , \\
            i J \left( 1_{\bar{Q}}^*, 2_Q^*, 3_g^+ \right) =  i &
            \frac{ |3] \langle n_3| \!+\! |n_3 \rangle [3| }
                 { \langle n_3 3 \rangle } .
      \label{M3}
	\end{align} \end{subequations}

      \begin{equation} \begin{aligned}
            i J \left( 1_{\bar{Q}}^*, 2_Q^{}, 3_g^-, 4_g^- \right)
            & = \frac{i}{ [n_3 3] [n_4 4] } \Bigg\{
                  \frac{1}{ \langle 3|2|3] }
                       \Bigg( |4 \rangle [n_4|2|3 \rangle [n_3|
                            - |n_4] \langle 4|1|n_3] \langle 3| \\
                        & + m |4 \rangle [n_4 n_3] \langle 3|
                          + m |n_4] \langle 4 3 \rangle [n_3| \Bigg)
                - \frac{1}{[34]}\Bigg( [n_4 3] \Big( |n_3] \langle 3| + |3 \rangle [n_3| \Big) \\
                                   & + [n_3 4] \Big( |n_4] \langle 4| + |4 \rangle [n_4| \Big)
                                     + \frac{[n_3 n_4]}{2}({\not}p_3 - {\not}p_4) \Bigg)
                \Bigg\} |2) .
      \label{iJ4mmq2}
	\end{aligned} \end{equation}

      \begin{equation} \begin{aligned}
            i J \left( 1_{\bar{Q}}^*, 2_Q^*, 3_g^-, 4_g^- \right)
            = - \frac{i}{ [n_3 3] [n_4 4] } \Bigg\{
              \Big( |n_4] \langle 4| \!+\! |4 \rangle [n_4| \Big)
              \frac{ {\not}p_2 \!-\! {\not}p_3 \!+\! m }{ (p_2 \!-\! p_3)^2 \!-\! m^2 }
            & \Big( |n_3] \langle 3| \!+\! |3 \rangle [n_3| \Big) \\
            + \frac{1}{[3 4]} \bigg[ \frac{[n_3 n_4]}{2}({\not}p_3 \!-\! {\not}p_4)
                                   + [n_3 4] \Big( |n_4] \langle 4| \!+\! |4 \rangle [n_4| \Big)
                                   + [n_4 3]&\Big( |n_3] \langle 3| \!+\! |3 \rangle [n_3| \Big)
                              \bigg]
              \Bigg\} ,
      \label{M4mm}
	\end{aligned} \end{equation}
      \begin{equation} \begin{aligned}
            i J \left( 1_{\bar{Q}}^*, 2_Q^*, 3_g^-, 4_g^+ \right)
            = \frac{i}{ [n_3 3] \langle n_4 4 \rangle } \Bigg\{
              \Big( |4] \langle n_4| \!+\! |n_4 \rangle [4| \Big)
              \frac{ {\not}p_2 \!-\! {\not}p_3 \!+\! m }{ (p_2 \!-\! p_3)^2 \!-\! m^2 }
            & \Big( |n_3] \langle 3| \!+\! |3 \rangle [n_3| \Big) \\
            + \frac{ [n_3 4] \langle n_4 3 \rangle }{ 2 [3 4] \langle 4 3 \rangle }
                                                       ({\not}p_3 \!-\! {\not}p_4)
            + \frac{ [n_3 4] }{ [3 4] } \Big( |4] \langle n_4| \!+\! |n_4 \rangle [4| \Big)
            - \frac{ \langle n_4 3 \rangle }{ \langle 4 3 \rangle }
                                      & \Big( |n_3] \langle 3| \!+\! |3 \rangle [n_3| \Big)
              \Bigg\} ,
	\end{aligned} \end{equation}
      \begin{equation} \begin{aligned}
            i J \left( 1_{\bar{Q}}^*, 2_Q^*, 3_g^+, 4_g^- \right)
            = \frac{i}{ \langle n_3 3 \rangle [n_4 4] } \Bigg\{
              \Big( |n_4] \langle 4| \!+\! |4 \rangle [n_4| \Big)
              \frac{ {\not}p_2 \!-\! {\not}p_3 \!+\! m }{ (p_2 \!-\! p_3)^2 \!-\! m^2 }
            & \Big( |3] \langle n_3| \!+\! |n_3 \rangle [3| \Big) \\
            + \frac{ \langle n_3 4 \rangle [n_4 3] }{ 2 \langle 3 4 \rangle [4 3] }
                                                       ({\not}p_3 \!-\! {\not}p_4)
            + \frac{ \langle n_3 4 \rangle }{ \langle 3 4 \rangle }
                                        \Big( |n_4] \langle 4| \!+\! |4 \rangle [n_4| \Big)
            - \frac{ [n_4 3] }{ [4 3] }&\Big( |3] \langle n_3| \!+\! |n_3 \rangle [3| \Big)
              \Bigg\} .
      \label{M4pm}
	\end{aligned} \end{equation}

\subsection{Closed form for all-minus currents}

For $n \geq 5$, we compute the all-minus current $i J \left( 1_{\bar{Q}}^*, 2_Q^{}, 3_g^-, 4_g^-,\ldots, n_g^- \right)$ by doing a $ [34 \rangle $ shift and setting all reference momenta equal to an arbitrary null vector $q$.  
Since we have established the absence of boundary terms and unphysical poles in the preceding sections, the BCFW expansion
is given as follows:
\bea
\label{bcfw-allminus}
i J \left( 1_{\bar{Q}}^*, 2_Q^{}, \hat{3}_g^-, \hat{4}_g^-,\ldots, n_g^- \right)
&=&
i J\left( 1_{\bar{Q}}^*, (2-\hat{3})_Q, \hat{4}_g^-,5_g^-,\ldots, n_g^- \right)
\frac{ i ( {\not}p_2 - {\not}\widehat{p}_3 + m ) }
     { (p_2 - {p}_3)^2 - m^2 }
i M \left(-(2-\hat{3})_{\bar{Q}} , 2_Q , \hat{3}_g^- \right)
\\ &&           
+ 
\sum_{(h,\tilde{h})}
\left[
i J\left( 1_{\bar{Q}}^*, 2_Q^{}, \hat{3}_g^-, (\hat{4}+5)_g^h,6_g^-,\ldots, n_g^- \right)
\frac{i}{ (p_4 + p_5)^2 }
            i M \left( -(\hat{4} \!+\!5)_g^{\tilde{h}}, \hat{4}_g^-, 5_g^- \right)
\nonumber \right. \\ && \left.
+\sum_{k=6}^n
i J\left( 1_{\bar{Q}}^*, 2_Q^{}, \hat{3}_g^-, (\hat{P}_{4,k})_g^h,(k+1)_g^-,\ldots, n_g^- \right)
\frac{i}{ P_{4,k}^2 }
            i M \left( -(\hat{P}_{4,k})_g^{\tilde{h}}, \hat{4}_g^-, 5_g^- ,\ldots,k_g^-\right)
\right]
\nonumber
\eea
      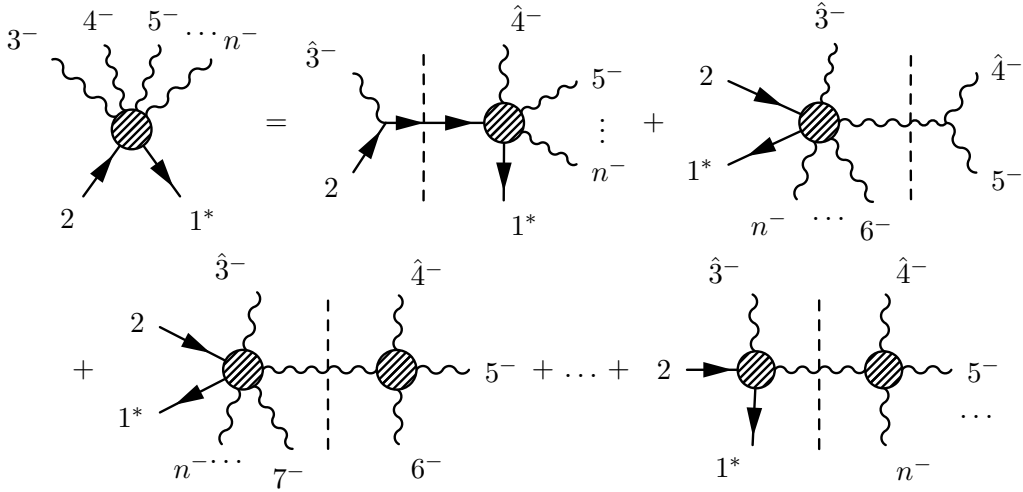
\begin{figure}
      \begin{center}
      \parbox{27mm}{ \begin{fmffile}{graph62}
      \fmfframe(0,12)(0,12){ \begin{fmfgraph*}(60,60)
            \fmflabel{$1^*$}{q1}
            \fmflabel{$2$}{q2}            
            \fmflabel{$3^-$}{g3}
            \fmflabel{$4^-$}{g4}
            \fmflabel{$5^-$}{g5}
            \fmflabel{$\dots$}{d}
            \fmflabel{$n^-$}{gn}
            \fmftop{g3,,,g4,,,g5,d,,gn}
            \fmfbottom{,q2,,,q1,}
            \fmf{fermion,tension=2.3}{q2,c,q1}
            \fmf{photon}{g3,c}
            \fmf{photon}{g4,c}
            \fmf{photon}{g5,c}
            \fmf{photon}{gn,c}
            \fmfblob{0.25h}{c}
      \end{fmfgraph*} }
      \end{fmffile} }
      $ = $ \hspace{5mm}
      \parbox{34mm}{ \begin{fmffile}{graph63}
      \fmfframe(0,12)(0,12){ \begin{fmfgraph*}(90,60)
            \fmflabel{$1^*$}{q1}
            \fmflabel{$2$}{q2}            
            \fmflabel{$\hat{3}^-$}{g3}
            \fmflabel{$\hat{4}^-$}{g4}
            \fmflabel{$5^-$}{g5}
            \fmflabel{$\vdots$}{d}
            \fmflabel{$n^-$}{gn}
            \fmfleft{,q2,,,g3,}
            \fmftop{,cut2,g4,}
            \fmfbottom{,cut1,q1,}
            \fmfright{,gn,d,g5,}
            \fmf{fermion}{q2,c3}
            \fmf{fermion,tension=1.7}{c3,cut}
            \fmf{fermion,tension=1.7}{cut,c5}
            \fmf{fermion,tension=10}{c5,q1}
            \fmf{photon}{g3,c3}
            \fmf{photon,tension=10}{g4,c5}
            \fmf{photon}{g5,c5}
            \fmf{photon}{gn,c5}
            \fmf{dashes,tension=100}{cut1,cut,cut2}
            \fmfblob{0.25h}{c5}
      \end{fmfgraph*} }
      \end{fmffile} }
      \hspace{3mm} $ + $ \hspace{5mm}
      \parbox{38mm}{ \begin{fmffile}{graph65}
      \fmfframe(0,12)(0,12){ \begin{fmfgraph*}(100,60)
            \fmflabel{$1^*$}{q1}
            \fmflabel{$2$}{q2}            
            \fmflabel{$\hat{3}^-$}{g3}
            \fmflabel{$\hat{4}^-$}{g4}
            \fmflabel{$5^-$}{g5}
            \fmflabel{$6^-$}{g6}
            \fmflabel{$\dots$}{d}
            \fmflabel{$n^-$}{gn}
            \fmfleft{,q1,,q2,}
            \fmftop{,,,g3,,cut2,,}
            \fmfbottom{,,gn,d,g6,cut1,,}
            \fmfright{,g5,,,g4,}
            \fmf{fermion,tension=3}{q2,c5,q1}
            \fmf{photon,tension=10}{g3,c5}
            \fmf{photon,tension=4}{gn,c5}
            \fmf{photon,tension=6}{g6,c5}
            \fmf{photon,tension=1.7}{cut,c5}
            \fmf{photon,tension=1.7}{cut,c3}
            \fmf{photon}{g4,c3}
            \fmf{photon}{g5,c3}
            \fmf{dashes,tension=100}{cut1,cut,cut2}
            \fmfblob{0.25h}{c5}
      \end{fmfgraph*} }
      \end{fmffile} } \\
      \vspace{3mm}
      $ + $ \hspace{5mm}
      \parbox{45mm}{ \begin{fmffile}{graph64}
      \fmfframe(0,12)(0,12){ \begin{fmfgraph*}(120,60)
            \fmflabel{$1^*$}{q1}
            \fmflabel{$2$}{q2}            
            \fmflabel{$\hat{3}^-$}{g3}
            \fmflabel{$\hat{4}^-$}{g4}
            \fmflabel{$5^-$}{g5}
            \fmflabel{$6^-$}{g6}
            \fmflabel{$7^-$}{g7}
            \fmflabel{$\dots$}{d}
            \fmflabel{$n^-$}{gn}
            \fmfleft{,q1,,q2,}
            \fmftop{,,,g3,,cut2,,g4,,}
            \fmfbottom{,,gn,d,g7,cut1,,g6,,}
            \fmfright{g5}
            \fmf{fermion,tension=3}{q2,c5,q1}
            \fmf{photon,tension=10}{g3,c5}
            \fmf{photon,tension=4}{gn,c5}
            \fmf{photon,tension=6}{g7,c5}
            \fmf{photon,tension=1.7}{cut,c5}
            \fmf{photon,tension=1.7}{cut,cg}
            \fmf{photon,tension=10}{g4,cg}
            \fmf{photon}{g5,cg}
            \fmf{photon,tension=10}{g6,cg}
            \fmf{dashes,tension=100}{cut1,cut,cut2}
            \fmfblob{0.25h}{c5}
            \fmfblob{0.25h}{cg}
      \end{fmfgraph*} }
      \end{fmffile} }
      \hspace{3mm} $ + \, \dots \, +  $ \hspace{5mm}
      \parbox{40mm}{ \begin{fmffile}{graph66}
      \fmfframe(0,12)(0,12){ \begin{fmfgraph*}(100,60)
            \fmflabel{$1^*$}{q1}
            \fmflabel{$2$}{q2}            
            \fmflabel{$\hat{3}^-$}{g3}
            \fmflabel{$\hat{4}^-$}{g4}
            \fmflabel{$5^-$}{g5}
            \fmflabel{$\dots$}{d}
            \fmflabel{$n^-$}{gn}
            \fmfleft{q2}
            \fmftop{,g3,cut2,g4,}
            \fmfbottom{,q1,cut1,gn,}
            \fmfright{,d,g5,,}
            \fmf{fermion,tension=0.5}{q2,c4}
            \fmf{fermion,tension=10}{c4,q1}
            \fmf{photon,tension=10}{g3,c4}
            \fmf{photon,tension=1.7}{cut,c4}
            \fmf{photon,tension=1.7}{cut,cg}
            \fmf{photon,tension=10}{g4,cg}
            \fmf{photon}{g5,cg}
            \fmf{photon,tension=10}{gn,cg}
            \fmf{dashes,tension=100}{cut1,cut,cut2}
            \fmfblob{0.14w}{c4}
            \fmfblob{0.14w}{cg}
      \end{fmfgraph*} }
      \end{fmffile} }
      \end{center}
      \caption{BCFW derivation of $i J \left( 1_{\bar{Q}}^*, 2_Q^{}, 3_g^-, 4_g^-,\ldots, n_g^- \right)$. \label{nmmmm}}
      \end{figure}
See Fig. \ref{nmmmm}.  
Because we are working with off-shell currents, 
the sum over intermediate gluon polarization states $(h,\tilde{h})$ must now include the unphysical polarization state combinations $(L,T), (T,L)$, which vanished automatically in the on-shell case due to the Ward identity.  This subtlety was first treated in a similar context in \cite{Feng:2011twa}.  Specifically, the numerator of the Feynman propagator is decomposed as 
      \begin{equation} \begin{aligned}
            -g_{\mu \nu} = \varepsilon_{\mu}^+ \varepsilon_{\nu}^-
                         + \varepsilon_{\mu}^- \varepsilon_{\nu}^+
                         + \varepsilon_{\mu}^L \varepsilon_{\nu}^T
                         + \varepsilon_{\mu}^T \varepsilon_{\nu}^L ,
      \label{metric}
	\end{aligned} \end{equation}
where 
      \begin{eqnarray} \begin{aligned}
            \varepsilon_{\mu}^L & = k_{\mu} ,  &
            \varepsilon_{\mu}^T & = - \frac{ q_{\mu} }{ kn } .
      \label{polvectorsfull}
	\end{aligned} \end{eqnarray}
	
It is clear that the gluon amplitude in the third line of (\ref{bcfw-allminus}) (the second line of Fig. \ref{nmmmm}) vanishes identically, due to the form of the all-minus Berends-Giele current (\ref{berendsgieleminus}) contracted with any of the polarization vectors.  The three-point gluon amplitude in the second line of (\ref{bcfw-allminus}) (the last diagram of the first line of Fig. \ref{nmmmm}) vanishes as well.  Therefore the only contribution that is left is the single term in the first line, involving a fermionic propagator.  The general expression for an $n$-point all-minus current can then be written as
      \begin{equation} \begin{aligned}
            | n^- (n\!-\!1)^- \dots 4^- 3^- | 2 ) = 
            & \frac{-i}{ [q \hat{3}] [q \hat{4}] \dots [q ~\widehat{n\!-\!2}] [q ~n\!-\!1] [q n] } \\
              \times \Bigg\{
              \Big( |q] \langle n| \!+\! |n \rangle [q| \Big)
            & \frac{ {\not}p_2 \!-\! {\not}\widehat{P}_{3,n\!-\!1} \!+\! m }
                   { (p_2 \!-\! P_{3,n\!-\!1})^2 \!-\! m^2 }
              \Big( |q] \langle \widehat{n\!-\!1}| \!+\! |\widehat{n\!-\!1} \rangle [q| \Big) \\
            + \frac{1}{[n\!-\!1~ n]} \bigg[ [q n]
              \Big( |q] \langle n| + & |n \rangle [q| \Big) + [q~n\!-\!1]
              \Big( |q] \langle \widehat{n\!-\!1}| \!+\! |\widehat{n\!-\!1} \rangle [q| \Big)
                                    \bigg] \Bigg\} \\ \times
            & \frac{ {\not}p_2 \!-\! {\not}\widehat{P}_{3,n\!-\!2} \!+\! m }
                   { (p_2 \!-\! P_{3,n\!-\!2})^2 \!-\! m^2 }
              \Big( |q] \langle \widehat{n\!-\!2}| \!+\! |\widehat{n\!-\!2} \rangle [q| \Big) \times \cdots \\ \times
            & \frac{ {\not}p_2 \!-\! {\not}\widehat{P}_{3,4} \!+\! m }
                   { (p_2 \!-\! P_{3,4})^2 \!-\! m^2 }
              \Big( |q] \langle \hat{4}| \!+\! |\hat{4} \rangle [q| \Big) \\ \times
            & \frac{ {\not}p_2 \!-\! {\not}\widehat{p}_3 \!+\! m }
                   { (p_2 \!-\! p_3)^2 \!-\! m^2 }
              \Big( |q] \langle 3| \!+\! |3 \rangle [q| \Big) |2) ,
      \label{iJminus}
	\end{aligned} \end{equation}
where the shifted momenta are defined recursively by
	\begin{equation}
      \left\{ \begin{aligned}
            z_k & = \frac{ \langle \hat{k}| {\not}p_2 \!-\! {\not}\widehat{P}_{3,k\!-\!1} |k] }
                         { \langle \hat{k}| {\not}p_2 \!-\! {\not}\widehat{P}_{3,k\!-\!1} |k\!+\!1] } \\
		|\hat{k}] & = |k] - z_k |k\!+\!1] \\
            |\widehat{k\!+\!1} \rangle & = |k\!+\!1 \rangle + z_k |\hat{k} \rangle \\
                   {\not}\widehat{P}_{3,k} &  \equiv {\not} P_{3,k} - z_k \Big( |k\!+\!1] \langle \hat{k}| \!+\! |\hat{k} \rangle [k\!+\!1| \Big) ,
	\end{aligned} \right.
      \label{shiftk}
	\end{equation}
with $ k = 3, 4, \dots, n - 2 $, and the initial values
\bea
 z_2=0, \qquad  z_3=\frac{\gb{3|2|3}}{\gb{3|2|4}}.
 \label{z-init}
\eea
 
 We have verified this formula numerically through $n=6$ by comparison with sums of Feynman diagrams.
 
 The massless version ($m=0$) was found in \cite{Berends:1987me} for one helicity choice of the on-shell spinor, namely $i J \left( 1_{\bar{Q}}^*, 2_Q^{+}, 3_g^-,  \ldots,  n_g^- \right)$ in our reversed fermion momentum convention, so that $|2)=\ket{2}$.  In our calculation, rather than take the massless limit of (\ref{iJminus}), it would be more effective to return to the recursion relation as given in the nonvanishing first line of (\ref{bcfw-allminus}), so that the propagators can be replaced by simple spinor products at each step of the recursion. Recovering the compact form of \cite{Berends:1987me} is not immediate for general $n$, however, because 
 we preserve a form of the current in which the quark spinor $|2)$ is an explicit factor at the right of the expression, free to take either helicity value. This is important, because the shift of the quark momentum means that the full internal helicity sum occurs at each stage of our recursion.
 
 It is possible to solve the recursion exactly and write the shifted spinors for (\ref{iJminus}) in a fully closed form.  We write this non-recursive form and outline its derivation in Appendix \ref{app:closedform}.

\subsection{Numerical results}

Beyond the case of all gluons having the same helicity, we have found valid shifts numerically through $n=6$ in the case of one gluon of opposite helicity to the others (the ``one-plus'' case or its parity conjugate).  A sufficient condition for a valid shift is to take the reference momenta of all the negative-helicity gluons equal to the momentum of the positive-helicity gluon: $n_- = p_+$.

For the choice of shifted gluons, we have identified two valid possibilities:
\begin{itemize}
      \item Shift the two gluons closest to the on-shell fermion; if they both have negative helicities, choose the shift so that the unphysical pole would come from the gluon adjacent to the on-shell fermion.  (These shifts are all valid in the all-minus case as well.)
      \item In the case with the plus-helicity gluon in central position shift the plus-gluon along with the any of the adjacent minus-gluons irrespective of their position with respect to the fermions. The unphysical poles then vanish, even with both fermions off-shell.
\end{itemize}

To be more precise, we found that for
$i J \left( 1_{\bar{Q}}^*, 2_Q^*, 3_g^-, 4_g^+, 5_g^-, 6_g^- \right)$
in gauge $ n_3 = n_5 = n_6 = p_4 $ not only $[34\rangle$ shift produces no unphysical poles (as we have proved in Section 4.3), but $[54\rangle$ as well. Similarly,
$i J \left( 1_{\bar{Q}}^*, 2_Q^*, 3_g^-, 4_g^-, 5_g^+, 6_g^- \right)$
in gauge $ n_3 = n_4 = n_6 = p_5 $ suffers from no unphysical poles both under $[65\rangle$ and $[45\rangle$ shifts.

      In the 6-point case we also have currents with two plus and two minus helicities, but unfortunately we were unable to find a good gauge choice for them.

\section{Summary and discussion}

We have studied currents of $n-2$ gluons of ``mostly-minus'' helicity and a massive quark-antiquark pair, where the antiquark is off shell.  
Because of the off-shellness of the antiquark, the choice of reference spinors plays an important role.
 
BCFW-type recursion relations are obtained under the following conditions, which ensure the absence of a boundary term and unphysical poles.
The reference spinors of the negative-helicity gluons are all chosen to be equal.  If there is a single positive-helicity gluon, its momentum is taken to be the reference spinor of the negative-helicity gluons.  
\begin{itemize}
\item In the case where all gluons have negative helicity, we have obtained both a recursive and a closed form for the current derived from recursion relations.

\item In the case where one gluon has positive helicity, and it is color-adjacent to the quark or antiquark, we have proven the validity of the recursion relation, but we do not have a closed form.

\item In the case where one gluon has positive helicity, and it is color-adjacent to two other gluons, we have found numerical evidence for the validity of the recursion relation in general, but were able to prove it only for the simplest configuration, with the positive gluon in next-to-extreme position.
\end{itemize}

In Yang-Mills theory, an on-shell alternative to the BCFW construction is the MHV diagram expansion \cite{Cachazo:2004kj},
in which maximally helicity violating (MHV) amplitudes play the role of interaction vertices, with a suitable on-shell prescription for the intermediate legs.
For off-shell currents, there is apparently no sensible expansion in MHV diagrams when the off-shell leg carries color charge, such as the Berends-Giele currents for gluons.

One might consider applying a BCFW shift to the massive fermion pair, but this construction fails off-shell.  
With a conventional definition of massive spinors \cite{Kleiss:1985yh,Schwinn:2005pi} in terms of a single reference vector, good boundary behavior is evident, but there are unavoidable, complicated unphysical poles, due to $\sqrt{z}$-dependence of the denominators of the massive spinors.
Even with both fermions on shell, the only shift known to be valid is quite specialized:  
each of the two massive fermion spinors has its reference vector constructed in terms of the other \cite{Boels:2011zz}.
This choice is not well suited for repeated application in an analytic recursion relation, because it is undesirable to keep track of the data of internal legs.  One would like the choice of reference spinor to be fixed once for all. Nevertheless, we looked at extending this construction off-shell.   There is no $z$-dependence in the denominators, but when either of the on-shell massive spinors are stripped off, the miraculous cancellation reducing the boundary behavior from $O(1)$ to $O(1/z)$ no longer takes place.

In the course of studying boundary behavior in Section 3, we have proven the good boundary behavior of general off-shell objects in Feynman gauge, as long as they contain at least two on-shell gluons that can be shifted. This meshes with a similar argument of \cite{Boels:2011mn} in the light-cone gauge $ q\!\cdot\! A = 0 $ specified by the BCFW-shift vector $q$, (\ref{q}). Thus we could see that it is not the boundary behaviour that hinders the BCFW recursion off shell, but the unphysical poles, coming from the polarization vectors.

Several questions arise for future exploration. Is there any choice of shift and reference spinors that eliminates boundary terms and unphysical poles for more general helicity configurations?  If so, can the recursion relation be solved neatly?  Do some shifts give more compact results than others?  Is there a neat solution for the current with a single gluon of opposite helicity, for which we have already proved the existence of recursion relations?  In cases where unphysical poles are present:  is there any way to understand them, so that their residues could be incorporated explicitly in the recursion relation?  Regarding the generalized Berends-Giele currents of Appendix \ref{app:berendsgiele}:  can the current with one opposite-helicity gluon in a central position be written in a compact form, manifestly independent of the generic reference spinor?  Further results addressing these questions would certainly illuminate our understanding of the BCFW construction and its applicability to gauge-dependent objects.

\section*{Acknowledgments}

We would like to thank Z. Bern, R. S. Isermann and E. Mirabella for helpful conversations.
This work is supported in part by the Agence Nationale de la Recherche under grant number ANR-09-CEXC-009-01, and by the Research Executive Agency (REA) of the European Union under the Grant Agreement number PITN-GA-2010-264564 (LHCPhenoNet).

\appendix

\section{Ward identity argument in Feynman gauge}
\label{app:wardid}

      In section 3.2, we used the Ward identity (\ref{ward}) to reduce the maximal superficial power of $z$ at infinity by two, but we omitted some extra terms that are present in Feynman gauge. In this appendix, we fill in this gap in our argument.

      The supplementary terms we need to consider are due to the fact that the Noether current of the global gauge transformation receives additional contributions from the gauge-fixing and ghost parts of the effective Lagrangian. (Note that in the generalized axial gauge the gauge-fixing term $ \propto \left( n^{\mu} A_{\mu}^a \right)^2 $ contains no derivatives and thus does not contribute to the Noether current.) At tree level, however, the ghost part of the current does not produce any non-vanishing diagrams, unless we consider a Green's function with external ghost legs, which is not the case. Thus the only non-trivial ingredient that we should worry about in Feynman gauge is the gauge-fixing contribution to the Noether current.

      To derive the Ward identities, we consider an infinitesimal global gauge transformation,
	\begin{equation}
            A_{\mu}^a \rightarrow A_\mu^a - g f^{abc} \alpha^b A_{\mu}^c ,
      \label{globalgauge}
	\end{equation}
which leaves invariant the $R_{\xi}$-gauge-fixing Lagrangian:
	\begin{equation}
            L_{\xi} = - \frac{1}{2 \xi} \left( \partial_{\mu} A^{a\mu} \right)^2 .
      \label{lagrangiangf}
	\end{equation}
The latter generates the following contribution to the Noether gauge current:
	\begin{equation}
            J_{\xi}^{a\mu} = \frac{\partial L_{\xi}}{\partial (\partial_{\mu} A_{\nu}^b)}
                             g f^{abc} A_{\nu}^c
                           = \frac{g}{\xi} f^{abc} A^{b\mu} \partial_{\nu} A^{c\nu} .
      \label{currentgf}
	\end{equation}
Taking a derivative gives
	\begin{equation*}
            \partial_\mu J_{\xi}^{a\mu}(x)
            = \frac{g}{\xi} f^{abc} A^{b\mu}(x) \partial_{\mu} \partial_{\nu} A^{c\nu}(x) ,
	\end{equation*}
so we retrieve the following momentum-space operator:
	\begin{equation}
            \hat{l}_\nu J_{\xi}^{b \nu} (\hat{l})
            = - \frac{i g}{\xi} f^{bcd} \int \frac{d^4 p}{(2\pi)^4}
                p_{\nu} p_{\rho} A^{c\nu}(p) A^{d\rho}(\hat{l}-p) .
      \label{ccurrentgf}
	\end{equation}
This operator is to be inserted instead of the $\hat{l}$-th leg and combined with the remaining $(n-1)$ legs of the off-shell current:
      \begin{equation}
      - \frac{i g}{\xi} f^{bcd} \int \frac{d^4 p}{(2\pi)^4}
                \hat{\varepsilon}_k^{\mu} p^{\nu} p^{\rho} 
      \left[ \parbox{55mm}{ \begin{fmffile}{graph32}
      \fmfframe(30,18)(0,18){ \begin{fmfgraph*}(80,60)
            \fmflabel{$1$}{q1}
            \fmflabel{$2$}{q2}
            \fmflabel{$3$}{g3}
            \fmflabel{$\vdots$}{d1}
            \fmflabel{$\hat{k},\mu,a$}{gk}
            \fmflabel{$\dots$}{d2}
            \fmflabel{$p,\nu,c$}{gp}
            \fmflabel{$\hat{l} \!-\! p,\rho,d$}{gl}
            \fmflabel{$\vdots$}{d3}
            \fmflabel{$n$}{gn}
            \fmfleft{q2,g3,,d1,,gk}
            \fmftop{d2}
            \fmfright{q1,gn,,d3,gl,gp}
            \fmf{fermion}{c,q1}
            \fmf{fermion}{q2,c}
            \fmf{photon}{g3,c}
            \fmf{photon}{gk,c}
            \fmf{photon,tension=0.5}{gp,c}
            \fmf{photon,tension=0.5}{gl,c}
            \fmf{photon}{gn,c}
            \fmfblob{0.25w}{c}
      \end{fmfgraph*} }
      \end{fmffile} } \right] .
      \label{wardextra}
      \end{equation}

      Note that in (\ref{wardextra}) only the $\hat{k}$-th leg is considered propagator-amputated, and we spell out its contraction with the polarization vector $\hat{\varepsilon}_k^{\mu}$ explicitly. In the following we specialize to the Feynman gauge $\xi = 1$ and once again neglect all color information. 
      Since (\ref{ccurrentgf}) has two gluon legs and already contains one power of $g$, then in order to construct a tree level contribution of order $O(g^{n-2})$ from an object with $(n+1)$ external legs we need to contract two of them together without any interaction insertions. Thus the other $(n-1)$ legs must form a normal connected tree level diagram of order $O(g^{n-3})$. An extra disconnected piece will naturally produce a $\delta$-function which will annihilate the integration in (\ref{wardextra}).

      We cannot contract together the two legs coming from (\ref{ccurrentgf}), because that would produce $ \delta^{(4)}(p+\hat{l}-p) = 0 $. Moreover, if both directly contracted legs are not in (\ref{ccurrentgf}), say the $i$-th and $j$-th, then we will get $ \delta^{(4)}(p_i+p_j) = 0 $. Obviously, we cannot contract directly a fermion with a gluon either. Thus we are left only with the options of connecting one of the two legs (\ref{ccurrentgf}) with any of the remaining gluon legs --- either the shifted one $\hat{k}$ or any of the unshifted legs.

      Contraction of the first leg of (\ref{ccurrentgf}) with the on-shell gluon $\hat{k}$ vanishes immediately due to the transversality of the polarization vector $\hat{\varepsilon}_{k-}^{\mu}$,
whereas connecting the other leg to $\hat{k}$ results in $ p = \hat{k} + \hat{l} $ :
      \begin{equation} \begin{aligned}
      & \hat{\varepsilon}_k^{\mu} p^{\nu} p^{\rho}
      \left[ \parbox{42mm}{
      \begin{fmffile}{graph35}
      \fmfframe(38,0)(0,-2){ \begin{fmfgraph*}(40,10)
            \fmflabel{$\hat{k},\mu$}{gk}
            \fmflabel{$\hat{l} \!-\! p,\rho$}{gl}
            \fmfleft{gk}
            \fmfright{gl}
            \fmf{photon}{gk,gl}
      \end{fmfgraph*} }
      \fmfframe(27,18)(0,18){ \begin{fmfgraph*}(70,50)
            \fmflabel{$1$}{q1}
            \fmflabel{$2$}{q2}
            \fmflabel{$3$}{g3}
            \fmflabel{$\vdots$}{d1}
            \fmflabel{$k\!-\!1$}{gk1}
            \fmflabel{$k\!+\!1$}{gk2}
            \fmflabel{$\dots$}{d2}
            \fmflabel{$p,\nu$}{gp}
            \fmflabel{$\vdots$}{d3}
            \fmflabel{$n$}{gn}
            \fmfleft{q2,g3,,d1,gk1,gk2}
            \fmftop{d2}
            \fmfright{q1,gn,,d3,,gp}
            \fmf{fermion}{c,q1}
            \fmf{fermion}{q2,c}
            \fmf{photon}{g3,c}
            \fmf{photon,tension=0.5}{gk1,c}
            \fmf{photon,tension=0.5}{gk2,c}
            \fmf{photon}{gp,c}
            \fmf{photon}{gn,c}
            \fmfblob{0.25w}{c}
      \end{fmfgraph*} }
      \end{fmffile} } \right] \\ & =
      (2\pi)^4 \delta^{(4)}(\hat{l}-p+\hat{k}) \hat{\varepsilon}_k^{\mu} (\hat{k}+\hat{l})_{\mu}
      \cdot (\hat{k}+\hat{l})^{\nu}
      \left[ \parbox{45mm}{
      \begin{fmffile}{graph36}
      \fmfframe(27,18)(0,18){ \begin{fmfgraph*}(70,50)
            \fmflabel{$1$}{q1}
            \fmflabel{$2$}{q2}
            \fmflabel{$3$}{g3}
            \fmflabel{$\vdots$}{d1}
            \fmflabel{$k\!-\!1$}{gk1}
            \fmflabel{$k\!+\!1$}{gk2}
            \fmflabel{$\dots$}{d2}
            \fmflabel{$\hat{k} \!+\! \hat{l},\nu$}{gp}
            \fmflabel{$\vdots$}{d3}
            \fmflabel{$n$}{gn}
            \fmfleft{q2,g3,,d1,gk1,gk2}
            \fmftop{d2}
            \fmfright{q1,gn,,d3,,gp}
            \fmf{fermion}{c,q1}
            \fmf{fermion}{q2,c}
            \fmf{photon}{g3,c}
            \fmf{photon,tension=0.5}{gk1,c}
            \fmf{photon,tension=0.5}{gk2,c}
            \fmf{photon}{gp,c}
            \fmf{photon}{gn,c}
            \fmfblob{0.25w}{c}
      \end{fmfgraph*} }
      \end{fmffile} } \right] = O\left( \frac{1}{z} \right) ,
      \label{wardextra2}
      \end{aligned} \end{equation}
so that the only remaining element dependent on $z$ is $ \hat{\varepsilon}_{k-}^{\mu} = O\left( \frac{1}{z} \right) $. Now if we take an arbitrary unshifted gluon leg with momentum $p_j$ and Lorentz index $\lambda$ and contract it with the leg $p$ from (\ref{ccurrentgf}), we get
      \begin{equation} \begin{aligned}
      & \hat{\varepsilon}_k^{\mu} p^{\nu} p^{\rho}
      \left[ \parbox{42mm}{
      \begin{fmffile}{graph37}
      \fmfframe(33,0)(0,-2){ \begin{fmfgraph*}(40,10)
            \fmflabel{$j,\lambda$}{gj}
            \fmflabel{$p,\nu$}{gp}
            \fmfleft{gj}
            \fmfright{gp}
            \fmf{photon}{gj,gp}
      \end{fmfgraph*} }
      \fmfframe(18,18)(0,18){ \begin{fmfgraph*}(70,50)
            \fmflabel{$1$}{q1}
            \fmflabel{$2$}{q2}
            \fmflabel{$3$}{g3}
            \fmflabel{$\vdots$}{d1}
            \fmflabel{$\hat{k},\mu$}{gk}
            \fmflabel{$\dots$}{d2}
            \fmflabel{$\hat{l}\!-\!p,\rho$}{gl}
            \fmflabel{$\vdots$}{d3}
            \fmflabel{$n$}{gn}
            \fmfleft{q2,g3,d1,gk}
            \fmftop{d2}
            \fmfright{q1,gn,d3,gl}
            \fmf{fermion}{c,q1}
            \fmf{fermion}{q2,c}
            \fmf{photon}{g3,c}
            \fmf{photon}{gk,c}
            \fmf{photon}{gl,c}
            \fmf{photon}{gn,c}
            \fmfblob{0.25w}{c}
      \end{fmfgraph*} }
      \end{fmffile} } \right] \\ & =
      (2\pi)^4 \delta^{(4)}(p+p_j) \frac{-i g_{\lambda \nu}}{p_j^2} p_j^{\nu}
      \cdot \hat{\varepsilon}_k^{\mu} p_j^{\rho}
      \left[ \parbox{43mm}{
      \begin{fmffile}{graph38}
      \fmfframe(18,18)(0,18){ \begin{fmfgraph*}(70,50)
            \fmflabel{$1$}{q1}
            \fmflabel{$2$}{q2}
            \fmflabel{$3$}{g3}
            \fmflabel{$\vdots$}{d1}
            \fmflabel{$\hat{k},\mu$}{gk}
            \fmflabel{$\dots$}{d2}
            \fmflabel{$\hat{l}\!+\!p_j,\rho$}{gl}
            \fmflabel{$\vdots$}{d3}
            \fmflabel{$n$}{gn}
            \fmfleft{q2,g3,d1,gk}
            \fmftop{d2}
            \fmfright{q1,gn,d3,gl}
            \fmf{fermion}{c,q1}
            \fmf{fermion}{q2,c}
            \fmf{photon}{g3,c}
            \fmf{photon}{gk,c}
            \fmf{photon}{gl,c}
            \fmf{photon}{gn,c}
            \fmfblob{0.25w}{c}
      \end{fmfgraph*} }
      \end{fmffile} } \right] \frac{-i}{(\hat{l}+p_j)^2}
      = O\left( \frac{1}{z} \right) .
      \label{wardextra3}
      \end{aligned} \end{equation}
In the second line we have written out the propagator of the $(\hat{l}-p)$ leg, making both $z$-dependent legs propagator-amputated, so that the diagram in the brackets is a standard $O(z)$ and the overall expression obviously vanishes at infinity.

      The last remaining type of contribution is not so straightforward. Connecting an off-shell gluon leg to the $(\hat{l}-p)$ leg makes $p$ equal to $\hat{l} + p_j$ and thus produces two more powers of $z$ in the numerator:
      \begin{equation} \begin{aligned}
      & \hat{\varepsilon}_k^{\mu} p^{\nu} p^{\rho}
      \left[ \parbox{40mm}{
      \begin{fmffile}{graph39}
      \fmfframe(33,0)(0,-2){ \begin{fmfgraph*}(40,10)
            \fmflabel{$j,\lambda$}{gj}
            \fmflabel{$\hat{l} \!-\! p,\rho$}{gl}
            \fmfleft{gj}
            \fmfright{gl}
            \fmf{photon}{gj,gl}
      \end{fmfgraph*} }
      \fmfframe(18,18)(0,18){ \begin{fmfgraph*}(70,50)
            \fmflabel{$1$}{q1}
            \fmflabel{$2$}{q2}
            \fmflabel{$3$}{g3}
            \fmflabel{$\vdots$}{d1}
            \fmflabel{$\hat{k},\mu$}{gk}
            \fmflabel{$\dots$}{d2}
            \fmflabel{$p,\nu$}{gp}
            \fmflabel{$\vdots$}{d3}
            \fmflabel{$n$}{gn}
            \fmfleft{q2,g3,d1,gk}
            \fmftop{d2}
            \fmfright{q1,gn,d3,gp}
            \fmf{fermion}{c,q1}
            \fmf{fermion}{q2,c}
            \fmf{photon}{g3,c}
            \fmf{photon}{gk,c}
            \fmf{photon}{gp,c}
            \fmf{photon}{gn,c}
            \fmfblob{0.25w}{c}
      \end{fmfgraph*} }
      \end{fmffile} } \right] \\ & =
      (2\pi)^4 \delta^{(4)}(\hat{l}-p+p_j) \frac{-i g_{\lambda \rho}}{p_j^2} (\hat{l}+p_j)^{\rho}
      \cdot \hat{\varepsilon}_k^{\mu} (\hat{l}+p_j)^{\nu}
      \left[ \parbox{43mm}{
      \begin{fmffile}{graph40}
      \fmfframe(18,18)(0,18){ \begin{fmfgraph*}(70,50)
            \fmflabel{$1$}{q1}
            \fmflabel{$2$}{q2}
            \fmflabel{$3$}{g3}
            \fmflabel{$\vdots$}{d1}
            \fmflabel{$\hat{k},\mu$}{gk}
            \fmflabel{$\dots$}{d2}
            \fmflabel{$\hat{l}\!+\!p_j,\nu$}{gp}
            \fmflabel{$\vdots$}{d3}
            \fmflabel{$n$}{gn}
            \fmfleft{q2,g3,d1,gk}
            \fmftop{d2}
            \fmfright{q1,gn,d3,gp}
            \fmf{fermion}{c,q1}
            \fmf{fermion}{q2,c}
            \fmf{photon}{g3,c}
            \fmf{photon}{gk,c}
            \fmf{photon}{gp,c}
            \fmf{photon}{gn,c}
            \fmfblob{0.25w}{c}
      \end{fmfgraph*} }
      \end{fmffile} } \right] \frac{-i}{(\hat{l}+p_j)^2} .
      \label{wardextra4}
      \end{aligned} \end{equation}
Fortunately, what we can see on the right-hand side apart from other $O(1)$ factors is just what we started with --- a BCFW-shifted off-shell current with both shifted legs propagator-amputated and contracted with the polarization vector $\hat{\varepsilon}_{k-}^{\mu}$ on one side and with the momentum $(\hat{l}+p_j)$ on the other. The main difference is that it has one gluon leg less than before and the momentum of the missing gluon is now added to the $\hat{l}$-th leg.

      So we started with a superficially $O(z)$ object with $(n-4)$ unshifted gluons on the left-hand side of (\ref{ward}), and we have managed to see that the right-hand side of the Ward identity gives us $O(\frac{1}{z})$ contributions plus $(n-4)$ superficially $O(z)$ objects of the same type, but with $(n-5)$ unshifted gluons. In the same fashion we can apply the Ward identity repeatedly until there are no unshifted gluons left, which proves that the initial object was indeed $O(\frac{1}{z})$.

      In this way we have verified that the Ward identity argument is still valid in Feynman gauge.

\section{Berends-Giele one-minus currents with a general reference spinor}
\label{app:berendsgiele}

Berends and Giele \cite{Berends:1987me} found 
a well-known recursion relation for multi-gluon currents.  
The explicit solutions they gave for currents with a single negative-helicity gluon require that it is color-adjacent to the off-shell leg, with a specific choice of reference momenta.
In this appendix, we relax these conditions, for use in the proofs of Section 4.

The Berends-Giele  solution for the current  $iJ^\mu(1^-,2^+,3^+,\ldots,n^+)$  assumed that the reference vectors $n_2,\ldots,n_n$ for the positive-helicity gluons were all equal to the momentum $p_1$, and that the reference vector $n_1$ was equal to $p_2$.  Under these conditions, all the polarization vectors are proportional to $\lambda_1$ and hence mutually orthogonal, and the Berends-Giele recursion simplifies considerably.  The orthogonality of polarization vectors is maintained even if we relax the second condition.  Keeping $n_2=n_3=\cdots=n_n=p_1$, but taking $n_1=q$, we find the following compact result:
\bea
i J^\mu(1^-,2^+,\ldots,n^+) &=&
\frac{\vev{1|\gamma^\mu {\not}P_{1,n}|1}}{\sqrt{2}\vev{12}\vev{23}\cdots\vev{n1}}
\left(
\frac{\cb{2q}}{\cb{21}\cb{1q}}+
\sum_{m=3}^n
\frac{\vev{1|{\not}P_{1,m}{\not}P_{1,m-1}|1}}{P_{1,m}^2 P_{1,m-1}^2}
\right).
 \label{berendsgielemostlyplus}
 \eea
 
 If the negative-helicity gluon is not color-adjacent to the off-shell line, we have a current of the form $i J^\mu(1^+,2^+,\ldots,m^-,\ldots,n^+)$.  If we choose reference vectors $n_1=n_2=\cdots=n_{m-1}=n_{m+1}=\cdots=n_n=p_m$ and $n_m=q$, we find by an inductive argument that the current is {\em independent} of $q$.  
 The base case with $n=3$ can be evaluated explicitly as
 \bea
 i J^\mu(1^+,2^-,3^+) = \frac{\vev{2|\gamma^\mu {\not}P_{1,3}|2}}{\sqrt{2}\vev{12}\vev{23}}
 \frac{\cb{13}^2}{\cb{12}\cb{23}}.
 \eea
 As in the all-plus case, all polarization vectors are proportional to $\lambda_m$ and hence orthogonal, giving the same simplified recursion (equation (5.47) of \cite{Berends:1987me}), which does not contain the four-point vertex.  
       \begin{figure}
      \begin{center}
      \parbox{40mm}{ \begin{fmffile}{graph60}
      \fmfframe(0,10)(0,10){ \begin{fmfgraph*}(75,75)
            \fmflabel{$\mu$}{mu}
            \fmflabel{$1^+$}{g1}
            \fmflabel{$2^+$}{g2}
            \fmflabel{$\ddots$}{d1}
            \fmflabel{$(m\!-\!1)^+$}{gm1}
            \fmflabel{$m^-$}{gm}
            \fmflabel{$(m\!+\!1)^+$}{gp1}
            \fmflabel{$\dots$}{d2}
            \fmflabel{$n^+$}{gn}
            \fmfleft{mu}
            \fmfbottom{,,,,gn,,d2,,}
            \fmfright{,gp1,gm,gm1,d1,}
            \fmftop{,,g1,g2,}
            \fmf{photon}{g1,c1}
            \fmf{photon}{g2,c1}
            \fmf{photon}{gm1,c1}
            \fmf{photon}{gm,c2}
            \fmf{photon}{gp1,c2}
            \fmf{photon}{gn,c2}
            \fmf{photon,tension=3}{c1,v,c2}
            \fmf{photon,tension=6}{mu,v}
            \fmfblob{0.20w}{c1}
            \fmfblob{0.20w}{c2}
      \end{fmfgraph*} }
      \end{fmffile} }
      \hspace{2mm} $ + $ \hspace{5mm}
      \parbox{40mm}{ \begin{fmffile}{graph61}
      \fmfframe(0,10)(0,10){ \begin{fmfgraph*}(75,75)
            \fmflabel{$\mu$}{mu}
            \fmflabel{$1^+$}{g1}
            \fmflabel{$\dots$}{d1}
            \fmflabel{$(m\!-\!1)^+$}{gm1}
            \fmflabel{$m^-$}{gm}
            \fmflabel{$(m\!+\!1)^+$}{gp1}
            \fmflabel{$(m\!+\!2)^+$}{gp2}
            \fmflabel{$\dots$}{d2}
            \fmflabel{$n^+$}{gn}
            \fmfleft{mu}
            \fmfbottom{,,,,gn,,d2,,}
            \fmfright{,gp2,gp1,gm,gm1,}
            \fmftop{,,,,g1,,d1,,}
            \fmf{photon}{g1,c1}
            \fmf{photon}{gm1,c1}
            \fmf{photon}{gm,c1}
            \fmf{photon}{gp1,c2}
            \fmf{photon}{gp2,c2}
            \fmf{photon}{gn,c2}
            \fmf{photon,tension=3}{c1,v,c2}
            \fmf{photon,tension=6}{mu,v}
            \fmfblob{0.20w}{c1}
            \fmfblob{0.20w}{c2}
      \end{fmfgraph*} }
      \end{fmffile} }
      \end{center}
      \caption{Berends-Giele derivation of $q$-independence of gluon current with a positive-helicity gluon in central position. \label{gaugecentralcurrent}}
      \end{figure}
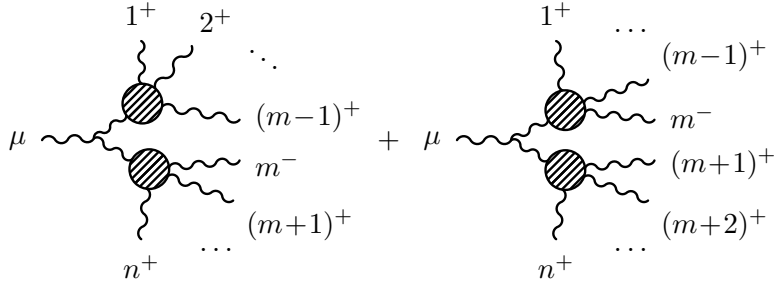
  According to the inductive assumption, 
it is enough to consider only the two contributions shown in Fig. \ref{gaugecentralcurrent}. Using the result (\ref{berendsgielemostlyplus}), we see that the $q$-dependent terms from the two diagrams combine into a $q$-independent contribution thanks to the Schouten identity.  It would be interesting to find a compact form for this current, preferably one that is manifestly independent of $q$.

\section{All-minus formula}
\label{app:closedform}

In this appendix, we present the shifted spinors for the current with all negative-helicity gluons, equation (\ref{iJminus}), in a fully closed, non-recursive form.  The shifted momenta are written as follows, in terms of shift parameters $z_k$ and new massless spinors $\sqket{t_k}$, both defined below.
	\begin{equation}
      \left\{ \begin{aligned}
		|\hat{k}] & = |t_k]\frac{\cb{k+1|k}}{\cb{k+1|t_k}} , \\
            |\widehat{k\!+\!1} \rangle & = \sum_{j=3}^{k+1} \ket{j} \left( \prod_{i=j}^k z_i \right) , \\
            {\not}\widehat{P}_{3,k} & \equiv  {\not} P_{3,k} - z_k \Big( |k\!+\!1] \langle \hat{k}| \!+\! |\hat{k} \rangle [k\!+\!1| \Big)      .
	\end{aligned} \right.
      \label{shiftk-closed}
	\end{equation}
The shift parameter is given by
\bea
           z_k =  \frac{\cb{t_k|k}}{\cb{t_k|k+1}}.
\label{zt-spinor}
\eea
The spinor $\sqket{t_k}$ is given by 
      \begin{equation}
            [t_k| = \left( \prod_{j=4}^{k} y_j \right) \langle 3|2|
                  + \sum_{j=4}^{k} \left( \prod_{i=j+1}^{k} y_i \right) [w_j| .
      \label{tspinor}
      \end{equation}
Here we use the  shorthand $y_k$ for the propagator denominator:
      \begin{equation}
            y_k = ( p_2 - P_{3,k} )^2 - m^2 ;
      \label{y}
      \end{equation}
the spinor $[w_k|$ takes the following values for $k=3,4$,
\bea
[w_3| = 0, \qquad [w_4| = m^2 \langle 3|4|; 
\eea
for $k \geq 5$,
      \begin{equation}
            [w_k| = [u_k| + [v_k| ,
      \label{wspinor}
      \end{equation}
where these elementary spinors $[u_k|$ and $[v_k|$ are calculated as follows:
      \begin{equation}
            [u_k| = - m^2 \sum_{p=0}^{[\frac{k-5}{2}]} \sum_{R_{ i_1 \dots i_p }^{4,k-2}}
            \left\{ \left(-m^2\right)^p \prod_{j\in S_{ i_1 \dots i_p }^{4,k-2}} y_j \right\}
            \langle 3|2| \prod_{ \substack{j = 1 \\ ( i_0 = 1 )} }
                              ^{ \substack{( i_{p+1} = k-1 ) \\ p+1 } }
                         \left[ {\not}P_{(i_{j-1}+2),i_j} \cdot
                                {\not}p_{(i_j+1)} \right] ,
      \label{uspinor}
      \end{equation}
      \begin{equation}
            [v_k| = - m^4 \sum_{p=0}^{[\frac{k-6}{2}]} \sum_{R_{ i_1 \dots i_p }^{5,k-2}}
            \left\{ \left(-m^2\right)^p \prod_{j\in S_{ i_1 \dots i_p }^{5,k-2}} y_j \right\}
            \langle 3|4| \prod_{ \substack{j = 1 \\ ( i_0 = 3 )} }
                              ^{ \substack{( i_{p+1} = k-1 ) \\ p+1 } }
                         \left[ {\not}P_{(i_{j-1}+2),i_j} \cdot
                                {\not}p_{(i_j+1)} \right].
      \label{vspinor}
      \end{equation}
The  sum is over all possible ordered pair lists $ R_{ i_1 \dots i_p }^{i,k} $, defined by
      \begin{equation}
            R_{ i_1 \dots i_p }^{i,k} =
                  \left\{ i_1, i_1 + 1;\ i_2, i_2 + 1;\ \dots\ ;\ i_p, i_p + 1 \right\} ,
      \label{pairs}
	\end{equation}
where $ i_1 \geq i $, $ i_p \leq k $, $ i_{j+1} \geq i_j+ 2 $; and the complement sets
      \begin{equation}
            S_{ i_1 \dots i_p }^{i,k} =
                  \left\{ i, i+1, \dots\, k \right\} \setminus R_{ i_1 \dots i_p }^{i,k} .
      \label{unpairs}
	\end{equation}
Finally, the products of ${\not}p$-matrices in (\ref{uspinor}) and (\ref{vspinor}) must be ordered by the ascending numbering of the gluon momenta.

\subsection{Derivation}

We now describe the derivation of the non-recursive 
 formula (\ref{shiftk-closed}). 
  
 The expressions for $|\hat{k}]$ and $\widehat{\ket{k+1}}$ follow directly from the recursions in (\ref{shiftk}), given the expression in (\ref{shiftk-closed}) for $z_k$ in terms of the new massless spinor $\sqket{t_k}$. It remains to justify this expression and the definitions of $\sqket{t_k}, \sqket{u_k}, \sqket{v_k}$.  
 
 Our derivation proceeds by induction on $k$.  We take $k=5$ as the base case.  Here, even though  the sets $R^{4,3}=S^{4,3} = \emptyset $, formula (\ref{uspinor}) already starts to produce a non-zero spinor $ [u_5| =  - m^2 \langle 3|2|{\not}P_{3,4}|5| $ from the index $p=0$, thanks to the boundary conditions $ i_0 = 1 $ and $ i_{p+1} = 4 $.  We have verified with a direct computation that this value produces the correct expression in (\ref{iJminus}).
 
 Assuming the validity of the formulas (\ref{zt-spinor}) -- (\ref{vspinor}) for $z_{k}, \sqbra{t_{k}}, \sqbra{u_{k}}, \sqbra{v_{k}}$, we now demonstrate their validity for $k \to k+1$.  By the BCFW construction, in which only the first line of equation (\ref{bcfw-allminus}) survives, $z_{k+1}$ is obtained from $z_k$ by making the following replacements of momenta, with the $[34\rangle$ shift:
      \begin{equation}
      \left\{ \begin{aligned}
             2 \:\, & \rightarrow \; 2 - \hat{3} \\
            |3 \rangle & \rightarrow | \hat{4} \rangle \\
            |3]\;\! & \rightarrow   |4] \\
             4 \:\, & \rightarrow \; 5 \\
             5 \:\, & \rightarrow \; 6 \\
            \dots \\
             k \:\, & \rightarrow \; k+1
	\end{aligned} \right.
      \label{replacement}
	\end{equation}
We denote the results of the replacement (\ref{replacement}) on $\sqbra{t_{k}}, \sqbra{u_{k}}, \sqbra{v_{k}}$ by $\sqbra{t'_{k}}, \sqbra{u'_{k}}, \sqbra{v'_{k}}$.
We now show that $\sqbra{t_{k+1}}$, defined by 
      \begin{equation}
            [t_{k+1}| = -\langle 3|2|4] [t_k'| = - \langle 3|2|4] \left\{
            \left( \prod_{j=5}^{k+1} y_j \right) \langle \hat{4} | 2\!-\!\hat{3} |
                  + \sum_{j=4}^{k} \left( \prod_{i=j+2}^{k+1} y_i \right) [w_j'| \right\} ,
      \label{zspinor1}
	\end{equation}
will once again satisfy the formulas (\ref{zt-spinor})--(\ref{vspinor}).  The prefactor  $-\langle 3|2|4]$ conveniently reproduces the formulas for $\sqbra{u_{k}}$ and  $\sqbra{v_{k}}$ and drops out of the ratio defining $z_{k+1}$.

Applying the relation  $- \langle 3|2|4] \langle \hat{4} | 2\!-\!\hat{3} | = y_4 \langle 3|2| + m^2 \langle 3|4| $
to the first term in (\ref{zspinor1}),
 we find
       \begin{equation} \begin{aligned}
            -\langle 3|2|4] \left( \prod_{j=5}^{k+1} y_j \right) \langle \hat{4} | 2\!-\!\hat{3} |
                  & = \left( \prod_{j=4}^{k+1} y_j \right) \langle 3|2|
                    + \left( \prod_{j=5}^{k+1} y_j \right) [w_4| .
      \label{firstterm}
	\end{aligned} \end{equation}
The second term in (\ref{zspinor1}) contains $ [w_j'|  = [u_j'| + [v_j'| $.  For the $[v_k'|$ term, we have
       \begin{equation} \begin{aligned}
            & -\langle 3|2|4] [v_k'| 
            =
            -m^4 \sum_{p=0}^{[\frac{k-6}{2}]} \sum_{R_{ i_1 \dots i_p }^{6,k-1}}
            \left\{ \left(-m^2\right)^p \prod_{j\in S_{ i_1 \dots i_p }^{6,k-1}} y_j \right\}
            \left\{ -\langle 3|2|4] \langle \hat{4} |5| \right\}
            \prod_{ \substack{j = 1 \\ ( i_0 = 4 )} }
                 ^{ \substack{( i_{p+1} = k ) \\ p+1 } }
                         \left[ {\not}P_{(i_{j-1}+2),i_j} \cdot
                                {\not}p_{(i_j+1)} \right] \\ & =
            +m^4 \sum_{p=0}^{[\frac{k-6}{2}]} \sum_{R_{ i_1 \dots i_p }^{6,k-1}}
            \left\{ \left(-m^2\right)^p \prod_{j\in S_{ i_1 \dots i_p }^{6,k-1}} y_j \right\}
            \langle 3|2|3\!+\!4|5|
            \prod_{ \substack{j = 1 \\ ( i_0 = 4 )} }
                 ^{ \substack{( i_{p+1} = k ) \\ p+1 } }
                         \left[ {\not}P_{(i_{j-1}+2),i_j} \cdot
                                {\not}p_{(i_j+1)} \right] \\ & =
            -m^2 \sum_{p=1}^{[\frac{k-4}{2}]} \sum_{R_{ i_1 \dots i_p }^{4,k-1}}
            \left\{ \left(-m^2\right)^p \prod_{j\in S_{ i_1 \dots i_p }^{4,k-1}} y_j \right\}
            \langle 3|2| \prod_{ \substack{j = 1 \\ ( i_0 = 1 )} }
                              ^{ \substack{( i_{p+1} = k ) \\ p+1 } }
                         \left[ {\not}P_{(i_{j-1}+2),i_j} \cdot
                                {\not}p_{(i_j+1)} \right] ,
      \label{vspinor1}
	\end{aligned} \end{equation}
where at the last step we included $|3\!+\!4|5|$ into the ordered product, so the last sum goes over only the pair sets of the form $ R_{ i_1 \dots i_p }^{4,k-1} = \{4,5\} \cup R_{ i_2 \dots i_p }^{6,k-1} $.

 For the $[u_k'|$ term, we start with
       \begin{equation} \begin{aligned}
            & -\langle 3|2|4] [u_k'| \\ & =
            -m^2 \sum_{p=0}^{[\frac{k-5}{2}]} \sum_{R_{ i_1 \dots i_p }^{5,k-1}}
            \left\{ \left(-m^2\right)^p \prod_{j\in S_{ i_1 \dots i_p }^{5,k-1}} y_j \right\}
            \left\{ -\langle 3|2|4] \langle \hat{4} | 2\!-\!\hat{3} | \right\}
            \prod_{ \substack{j = 1 \\ ( i_0 = 2 )} }
                 ^{ \substack{( i_{p+1} = k ) \\ p+1 } }
                         \left[ {\not}\widehat{P}_{(i_{j-1}+2),i_j} \cdot
                                {\not}p_{(i_j+1)} \right],
      \label{uspinor1}
	\end{aligned} \end{equation}
where the hat over $ {\not}\widehat{P}_{(i_{j-1}+2),i_j} $ indicates only that the ordered product of ${\not}p~$'s starts with $ | \hat{4} \!+\! 5 \dots | $.
 We then apply the relation
  $ - \langle 3|2|4] \langle \hat{4} | 2\!-\!\hat{3} | \hat{4}\!+\!5\! +\!\cdots |
                  = y_4 \langle 3|2| 3\!+\!4\!+\!5 +\!\cdots | + m^2 \langle 3|4|5+\!\cdots | $
to separate the terms in (\ref{uspinor1}) into two types: those starting with 
$ \langle 3|2| 3\!+\!4\!+\!5 \dots | $ and those starting with $ \langle 3|4|5 \dots| $. Both can be easily incorporated into the ordered products just by shifting $i_0$, giving 
     \begin{equation} \begin{aligned}
            -\langle 3|2|4] [u_k'| 
            =
            -m^2 & \sum_{p=0}^{[\frac{k-5}{2}]} \sum_{R_{ i_1 \dots i_p }^{5,k-1}}
              \left\{ \left(-m^2\right)^p \prod_{j\in S_{ i_1 \dots i_p }^{4,k-1}} y_j \right\}
              \langle 3|2| \prod_{ \substack{j = 1 \\ ( i_0 = 1 )} }
                                ^{ \substack{( i_{p+1} = k ) \\ p+1 } }
                           \left[ {\not}P_{(i_{j-1}+2),i_j} \cdot
                                  {\not}p_{(i_j+1)} \right] \\
            -m^4 & \sum_{p=0}^{[\frac{k-5}{2}]} \sum_{R_{ i_1 \dots i_p }^{5,k-1}}
              \left\{ \left(-m^2\right)^p \prod_{j\in S_{ i_1 \dots i_p }^{5,k-1}} y_j \right\}
              \langle 3|4| \prod_{ \substack{j = 1 \\ ( i_0 = 3 )} }
                                ^{ \substack{( i_{p+1} = k ) \\ p+1 } }
                           \left[ {\not}P_{(i_{j-1}+2),i_j} \cdot
                                  {\not}p_{(i_j+1)} \right] ,
      \label{uspinor1s}
	\end{aligned} \end{equation}
where we have added $\{4\}$ to $ S_{ i_1 \dots i_p }^{4,k-1} = \{4\} \cup S_{ i_1 \dots i_p }^{5,k-1} $ by hand in order to include $y_4$ into the coefficient of the first term. The last term in (\ref{uspinor1}) is precisely $[v_{k+1}|$.

Summing (\ref{vspinor1}) and (\ref{uspinor1s}), we obtain
      \begin{equation}
            -\langle 3|2|4] [w_k'| = [u_{k+1}| + [v_{k+1}| = [w_{k+1}|.
      \label{wspinor1}
	\end{equation}
      Finally, plugging (\ref{firstterm}) and (\ref{wspinor1}) into (\ref{zspinor1}) we retrieve
      \begin{equation} \begin{aligned}
            -\langle 3|2|4] [z_k'| 
                & = \left( \prod_{j=4}^{k+1} y_j \right) \langle 3|2|
                  + \sum_{j=4}^{k+1} \left( \prod_{i=j+1}^{k+1} y_i \right) [w_j|,
      \label{zspinor1s}
	\end{aligned} \end{equation}
which is exactly our formula for $[z_{k+1}|$.

\bibliographystyle{JHEP}
\bibliography{references}

\providecommand{\href}[2]{#2}\begingroup\raggedright\begin{thebibliography}{10}

\bibitem{Berends:1987me}
F.~A. Berends and W.~Giele, {\it {Recursive Calculations for Processes with n
  Gluons}},  {\em Nucl.Phys.} {\bf B306} (1988) 759.

\bibitem{Kosower:1989xy}
D.~A. Kosower, {\it {Light Cone Recurrence Relations for QCD Amplitudes}},
  {\em Nucl.Phys.} {\bf B335} (1990) 23.

\bibitem{Feng:2011twa}
B.~Feng and Z.~Zhang, {\it {Boundary Contributions Using Fermion Pair
  Deformation}},  {\em JHEP} {\bf 1112} (2011) 057,
  [\href{http://xxx.lanl.gov/abs/1109.1887}{{\tt arXiv:1109.1887}}].

\bibitem{Britto:2011cr}
R.~Britto and E.~Mirabella, {\it {External leg corrections in the unitarity
  method}},  {\em JHEP} {\bf 1201} (2012) 045,
  [\href{http://xxx.lanl.gov/abs/1109.5106}{{\tt arXiv:1109.5106}}].

\bibitem{Rodrigo:2005eu}
G.~Rodrigo, {\it {Multigluonic scattering amplitudes of heavy quarks}},  {\em
  JHEP} {\bf 0509} (2005) 079,
  [\href{http://xxx.lanl.gov/abs/hep-ph/0508138}{{\tt hep-ph/0508138}}].

\bibitem{Britto:2005fq}
R.~Britto, F.~Cachazo, B.~Feng, and E.~Witten, {\it {Direct proof of tree-level
  recursion relation in Yang-Mills theory}},  {\em Phys.Rev.Lett.} {\bf 94}
  (2005) 181602, [\href{http://xxx.lanl.gov/abs/hep-th/0501052}{{\tt
  hep-th/0501052}}].

\bibitem{Badger:2005jv}
S.~Badger, E.~N. Glover, and V.~V. Khoze, {\it {Recursion relations for gauge
  theory amplitudes with massive vector bosons and fermions}},  {\em JHEP} {\bf
  0601} (2006) 066, [\href{http://xxx.lanl.gov/abs/hep-th/0507161}{{\tt
  hep-th/0507161}}].

\bibitem{Feng:2009ei}
B.~Feng, J.~Wang, Y.~Wang, and Z.~Zhang, {\it {BCFW Recursion Relation with
  Nonzero Boundary Contribution}},  {\em JHEP} {\bf 1001} (2010) 019,
  [\href{http://xxx.lanl.gov/abs/0911.0301}{{\tt arXiv:0911.0301}}].

\bibitem{Feng:2010ku}
B.~Feng and C.-Y. Liu, {\it {A Note on the boundary contribution with bad
  deformation in gauge theory}},  {\em JHEP} {\bf 1007} (2010) 093,
  [\href{http://xxx.lanl.gov/abs/1004.1282}{{\tt arXiv:1004.1282}}].

\bibitem{Benincasa:2011kn}
P.~Benincasa and E.~Conde, {\it {On the Tree-Level Structure of Scattering
  Amplitudes of Massless Particles}},  {\em JHEP} {\bf 1111} (2011) 074,
  [\href{http://xxx.lanl.gov/abs/1106.0166}{{\tt arXiv:1106.0166}}].

\bibitem{Benincasa:2011pg}
P.~Benincasa and E.~Conde, {\it {Exploring the S-Matrix of Massless
  Particles}},  {\em Phys.Rev.} {\bf D86} (2012) 025007,
  [\href{http://xxx.lanl.gov/abs/1108.3078}{{\tt arXiv:1108.3078}}].

\bibitem{Quigley:2005cu}
C.~Quigley and M.~Rozali, {\it {Recursion relations, helicity amplitudes and
  dimensional regularization}},  {\em JHEP} {\bf 0603} (2006) 004,
  [\href{http://xxx.lanl.gov/abs/hep-ph/0510148}{{\tt hep-ph/0510148}}].

\bibitem{Schwinn:2006ca}
C.~Schwinn and S.~Weinzierl, {\it {SUSY ward identities for multi-gluon
  helicity amplitudes with massive quarks}},  {\em JHEP} {\bf 0603} (2006) 030,
  [\href{http://xxx.lanl.gov/abs/hep-th/0602012}{{\tt hep-th/0602012}}].

\bibitem{Ozeren:2006ft}
K.~Ozeren and W.~Stirling, {\it {Scattering amplitudes with massive fermions
  using BCFW recursion}},  {\em Eur.Phys.J.} {\bf C48} (2006) 159--168,
  [\href{http://xxx.lanl.gov/abs/hep-ph/0603071}{{\tt hep-ph/0603071}}].

\bibitem{Ferrario:2006np}
P.~Ferrario, G.~Rodrigo, and P.~Talavera, {\it {Compact multigluonic scattering
  amplitudes with heavy scalars and fermions}},  {\em Phys.Rev.Lett.} {\bf 96}
  (2006) 182001, [\href{http://xxx.lanl.gov/abs/hep-th/0602043}{{\tt
  hep-th/0602043}}].

\bibitem{Schwinn:2007ee}
C.~Schwinn and S.~Weinzierl, {\it {On-shell recursion relations for all Born
  QCD amplitudes}},  {\em JHEP} {\bf 0704} (2007) 072,
  [\href{http://xxx.lanl.gov/abs/hep-ph/0703021}{{\tt hep-ph/0703021}}].

\bibitem{Hall:2007mz}
A.~Hall, {\it {Massive Quark-Gluon Scattering Amplitudes at Tree Level}},  {\em
  Phys.Rev.} {\bf D77} (2008) 025011,
  [\href{http://xxx.lanl.gov/abs/0710.1300}{{\tt arXiv:0710.1300}}].

\bibitem{Chen:2011sba}
G.~Chen, {\it {Recursion relations for the general tree-level amplitudes in QCD
  with massive dirac fields}},  {\em Phys.Rev.} {\bf D83} (2011) 125005,
  [\href{http://xxx.lanl.gov/abs/1103.2518}{{\tt arXiv:1103.2518}}].

\bibitem{Boels:2011zz}
R.~H. Boels and C.~Schwinn, {\it {On-shell supersymmetry for massive
  multiplets}},  {\em Phys.Rev.} {\bf D84} (2011) 065006,
  [\href{http://xxx.lanl.gov/abs/1104.2280}{{\tt arXiv:1104.2280}}].

\bibitem{Huang:2012gs}
J.-H. Huang, W.~Wang, and W.~Wang, {\it {Multigluon tree amplitudes with a pair
  of massive fermions}},  {\em Eur.Phys.J.} {\bf C72} (2012) 2050,
  [\href{http://xxx.lanl.gov/abs/1204.0068}{{\tt arXiv:1204.0068}}].

\bibitem{Berends:1981rb}
F.~A. Berends, R.~Kleiss, P.~De~Causmaecker, R.~Gastmans, and T.~T. Wu, {\it
  {Single Bremsstrahlung Processes in Gauge Theories}},  {\em Phys. Lett.} {\bf
  B103} (1981) 124.

\bibitem{DeCausmaecker:1981bg}
P.~De~Causmaecker, R.~Gastmans, W.~Troost, and T.~T. Wu, {\it {Multiple
  Bremsstrahlung in Gauge Theories at High- Energies. 1. General Formalism for
  Quantum Electrodynamics}},  {\em Nucl. Phys.} {\bf B206} (1982) 53.

\bibitem{Kleiss:1985yh}
R.~Kleiss and W.~J. Stirling, {\it {Spinor Techniques for Calculating $p \bar p
  \to W^\pm / Z^0$ + Jets}},  {\em Nucl. Phys.} {\bf B262} (1985) 235--262.

\bibitem{Xu:1986xb}
Z.~Xu, D.-H. Zhang, and L.~Chang, {\it {Helicity Amplitudes for Multiple
  Bremsstrahlung in Massless Nonabelian Gauge Theories}},  {\em Nucl. Phys.}
  {\bf B291} (1987) 392.

\bibitem{Gunion:1985vca}
J.~F. Gunion and Z.~Kunszt, {\it {Improved Analytic Techniques for Tree Graph
  Calculations and the G g q anti-q Lepton anti-Lepton Subprocess}},  {\em
  Phys. Lett.} {\bf B161} (1985) 333.

\bibitem{Schwinn:2005pi}
C.~Schwinn and S.~Weinzierl, {\it {Scalar diagrammatic rules for Born
  amplitudes in QCD}},  {\em JHEP} {\bf 0505} (2005) 006,
  [\href{http://xxx.lanl.gov/abs/hep-th/0503015}{{\tt hep-th/0503015}}].

\bibitem{Cachazo:2004kj}
F.~Cachazo, P.~Svrcek, and E.~Witten, {\it {MHV vertices and tree amplitudes in
  gauge theory}},  {\em JHEP} {\bf 09} (2004) 006,
  [\href{http://xxx.lanl.gov/abs/hep-th/0403047}{{\tt hep-th/0403047}}].

\bibitem{Boels:2011mn}
R.~H. Boels and R.~S. Isermann, {\it {Yang-Mills amplitude relations at loop
  level from non-adjacent BCFW shifts}},  {\em JHEP} {\bf 1203} (2012) 051,
  [\href{http://xxx.lanl.gov/abs/1110.4462}{{\tt arXiv:1110.4462}}].

\end{thebibliography}\endgroup

\end{document}